\documentclass[
aip,
amsmath,amssymb,preprint,floatfix
]{revtex4-1}

\usepackage{dcolumn}
\usepackage{bm}
\usepackage{physics}
\usepackage{verbatim}
\usepackage{hyperref}
\usepackage{natbib}
\usepackage{placeins}
\usepackage{adjustbox}
\usepackage{tikz}
\usepackage{soul}
\usepackage{subcaption}
\usepackage{cleveref}
\usepackage{graphicx}
\usepackage{float}
\usepackage{xcolor}
\usepackage{mathtools}
\renewcommand{\color}[2][]{}

\usepackage{soul} 

\Crefname{equation}{Eq.}{Eqs.}
\Crefname{figure}{Fig.}{Figs.}

 \definecolor{red}{rgb}{1.0,0.0,0.0} 
 \definecolor{gre}{rgb}{0.5,0.5,1.5}
 \definecolor{blu}{rgb}{0.0,0.0,1.0}

\begin{document}

\title{Near-ideal relaxed MHD in slab geometry}
\author{Arash Tavassoli}
    \affiliation{Mathematical Sciences Institute, Australian National University, Acton, ACT 2601, Australia}
	\email{arash.tavassoli@anu.edu.au}
 \author{Stuart R. Hudson}
 \affiliation{Princeton Plasma Physics Laboratory, PO Box 451, Princeton NJ 08543, USA.}
 \email{shudson@pppl.gov}
  \author{Zhisong Qu}
 \affiliation{School of Physical and Mathematical Sciences, Nanyang Technological University,  Singapore 637371, Singapore}
 \author{Matthew Hole}
     \affiliation{Mathematical Sciences Institute, Australian National University, Acton, ACT 2601, Australia}

\begin{abstract}
We investigate the solutions of the relaxed MHD model (RxMHD) of Dewar \& Qu [J. Plasma Phys. {\bf 88}, 835880101 (2022)]. 
This model generalizes Taylor relaxation by including the ideal Ohm's law constraint using an augmented Lagrangian method, providing a pathway to extend the multi-region relaxed MHD (MRxMHD) model.
We present the first numerical solution of the RxMHD model by Dewar \& Qu, demonstrating that it is mathematically well-defined and computationally feasible for constructing MHD equilibria in slab geometry.
We also show that a cross-field flow can exist without enforcing an arbitrary constraint on the angular momentum, as is done in the case of MRxMHD with flow.
Our results also demonstrate the self-organization of fully relaxed regions during the optimization, which was an important motivation behind developing this model.
\end{abstract}
 \maketitle

 \section{Introduction}
 A central challenge in magnetic confinement is constructing a magnetohydrodynamic (MHD) equilibrium that is sufficiently smooth, global, and consistent with the underlying MHD physics.
 A pioneering equilibrium model was proposed by Kruskal and Kulsrud\cite{kruskal1958equilibrium}. 
 In this model, the magnetic field is constrained to be tangential to a continuous set of nested flux surfaces, which act as transport barriers and are ergodically covered by magnetic field lines. Therefore, varying the geometry of these surfaces, which from a Hamiltonian mechanical viewpoint are equivalent to the Kolmogorov-Arnold-Moser (KAM) tori\cite{lichtenberg2013regular}, is equivalent to varying the magnetic field.
 The problem is then reduced to an optimization problem that minimizes a Hamiltonian functional, subject to constraints of ideal MHD. 
 During the optimization, the geometry of magnetic flux surfaces can evolve, while the pressure remains constant on each surface. A well-known problem of this model is that the frozen-in flux constraint of ideal MHD prevents topological changes, 
 magnetic reconnection is excluded and singularities (in the parallel current density) can arise in the solutions\cite{rodriguez2021islands,hudson2010pressure}.
Another issue is that the optimization model is constrained by an infinite set of ideal MHD constraints, and its solutions are highly sensitive to the initial conditions. As a result, these solutions fail to meet a key characteristic of equilibrium: they should be largely independent of the initial conditions, i.e.~they should be ``global"\cite{taylor1986relaxation}. 
To address these issues, Taylor suggests a heuristic model, known as ``Taylor relaxation", in which from the infinity of constraints of ideal MHD only the conservation of total magnetic helicity is held \cite{taylor1974relaxation}. The pressure gradient is zero and the magnetic field satisfies the Beltrami equation with a constant multiplier, which can accommodate complicated field lines--phenomena such as islands and chaos. Taylor relaxation has successfully described various phenomena, mainly in the reversed field pinch configuration and spheromaks \cite{taylor1986relaxation,bodin1990reversed,yamada2010magnetic}. 
Nevertheless, Taylor relaxation is too simple to describe the equilibrium in some fusion devices such as tokamaks and stellarators, where a non-constant pressure profile typically peaks at the magnetic axis and decreases towards the plasma edge.

This shortcoming is addressed in Multi-Region Relaxed MHD (MRxMHD)\cite{hole2006stepped,hudson2012computation}.  
In this model, the nested flux surfaces are discretely positioned in the radial direction \textcolor{blue}{(the direction perpendicular to the flux surfaces)} and these support all of the pressure gradient. 
In response, the flux surfaces (also called ideal interfaces) act as $\delta$-function current sheets which exert an electromagnetic force that balances the exerted pressure gradient \cite{dewar2008relaxed}. 
While the Kruskal--Kulsrud model continuously constrains the geometry, the MRxMHD discretely constrains the geometry.
Similar to the Kruskal--Kulsrud model, in MRxMHD, the flux surfaces are assumed to remain resilient against magnetic reconnection while their geometry is optimized to minimize the Hamiltonian functional. 
In the (sub-)volume enclosed by two consecutive flux surfaces, the pressure is constant and the Taylor relaxation allows the field lines to reconnect and islands to form. The Stepped Pressure Equilibrium Code (SPEC) works based on the same model \cite{hudson2012computation}. \textcolor{blue}{SPEC also offers the flexibility to replace the helicity constraint of the Taylor relaxation with either a rotational transform constraint or a Beltrami's constant constraint. The latter is employed to prescribe the parallel current profile instead of the helicity profile, because, during slow magnetic reconnection (on a timescale much longer than the Alfvén time), the conserved quantity is expected to be the parallel current rather than the helicity.\cite{loizu2020direct}.}
Several studies have confirmed the ability of SPEC to construct 3-dimensional (3D) equilibria\cite{dennis2013minimally,loizu2017equilibrium,loizu2016verification,hudson2020free} and even to predict the equilibrium state of tearing instabilities\cite{loizu2020direct,loizu2023nonlinear,balkovic2024direct}. 
Newer versions of SPEC can also accommodate the equilibrium with flow \cite{qu2020stepped,dennis2014multi}, with cross-magnetic field flow added by an angular momentum constraint.
However, this constraint (and therefore the resultant cross-field flow) is only applicable in sub-volumes that are bounded by axisymmetric interfaces. \textcolor{blue}{This condition limits the inclusion of cross-field flow in SPEC via the angular momentum constraint to only a few specific cases.}

SPEC-like sharp boundary equilibria with prescribed pressure profiles are supported by the KAM theorem and Bruno and Lawrence theorem \cite{bruno1996existence} for the systems whose departure from axisymmetry is small.
However, neither of these theorems ensures the existence of such equilibria in the presence of arbitrary perturbations. Ref.~\onlinecite{qu2021non} shows that enforcing indestructible flux surfaces in fixed radial positions along with a stepped pressure profile, as is done in SPEC, might lead to the nonexistence of a solution, \textcolor{blue}{which is evident in the failure of the numerical solution to converge as the number of Fourier components increases.} This happens when the enforced position of flux surfaces coincides with the chaotic field lines of the corresponding single-volume Taylor relaxed state or when the enforced pressure jumps are too large. This problem might be rooted in the fact that in SPEC, the existence of a certain number of (Taylor) relaxed regions is postulated.  Another potential problem of MRxMHD is the Taylor relaxation in each volume. Several studies suggest that besides magnetic helicity, other constraints can remain conserved during
a magnetic reconnection event\cite{bhattacharjee1980energy,yeates2010topological,amari2000helicity}. \textcolor{blue}{Therefore, there is motivation to develop an extended MRxMHD model that potentially mitigates the aforementioned issues of the MRxMHD.}

Ref.~\onlinecite{dewar2015variational} develops the variational formulation of the MRxMHD and lays the groundwork for the extension of Taylor relaxation based on Hamilton's action principle. 
Following this principle, Ref.~\onlinecite{dewar2020time} proposes Relaxed MHD (RxMHD) as an extension of single-region Taylor relaxation to a time-dependent problem that includes entropy and cross-helicity constraints as well as the helicity constraint. 
The basis of RxMHD is the ``Ideal MHD Consistency Principle", meaning solutions of RxMHD are a subset of all possible solutions of the ideal MHD. 
This principle is guaranteed if the constraints of RxMHD are a subset of the constraints of ideal MHD. 
Ref.~\onlinecite{dewar2020time} uses a {\it phase space Lagrangian} instead of a configuration space Lagrangian, which allows for defining two flows; one is the Lagrangian flow which can have both perpendicular and parallel components, and the other one is an exclusively field-aligned flow that sets the cross-helicity constraint. 
The total flow, defined by the momentum density, is the sum of the two flows and is shown to satisfy the equation of motion of the ideal MHD. 
Therefore, unlike MRxMHD, the cross-field flow can exist regardless of the axisymmetry attribute of the enclosing boundaries. 
By introducing a weak form of the ideal Ohm's law (IOL), Dewar and Qu\cite{dewar2022relaxed} make this model more consistent with the ideal MHD.  
Using the augmented Lagrangian method and in several steps, the model is brought closer to the ideal MHD while it does not exactly enforce the IOL. 
The final model aims to be (at least locally) consistent with ideal MHD, while it is relaxed to an extent just enough to accommodate magnetic reconnection and islands. Therefore, the work of Dewar and Qu\cite{dewar2022relaxed} can be viewed as a ``regularization" of the ideal MHD. This model proposes a pathway for the development of an extended MRxMHD, where the Taylor relaxation in each volume is replaced by RxMHD with a weak IOL constraint. Unlike MRxMHD this model allows for the self-organization of fully relaxed regions during the optimization process, resulting in a more physically reliable approach. \textcolor{blue}{Additionally, R.L. Dewar conjectured that the new model might help mitigate the non-existence issues of MRxMHD \cite{DewarPrivateComm}.}

In this work, we demonstrate the first nonlinear numerical solution of the relaxed MHD model proposed by Dewar and Qu \cite{dewar2022relaxed}, under the assumption of equilibrium in slab geometry. 
\textcolor{blue}{We show that,  within the specified assumptions, the proposed model is well-defined and converges.} This work aims to set the stage for developing a numerical code of ``extended MRxMHD model", based on the RxMHD with the IOL constraint. 

\section{Setup of the equilibrium problem as a Hamiltonian minimization problem}\label{sec:setup}
\textcolor{blue}{In Ref.~\onlinecite{dewar2022relaxed}, two tasks have been prescribed for future studies. ``Task 1" is prescribed as an equilibrium problem, which focuses on the Hamiltonian minimization (Eq.~(4.2) in Ref.~\onlinecite{dewar2022relaxed}). ``Task 2" is the extension of Task 1 to the time-dependent problem (Eq.~(4.8) in Ref.~\onlinecite{dewar2022relaxed}). Here, we undertake the Task 1, and therefore we find the stationary points of the Hamiltonian functional.}

The Hamiltonian that one needs to minimize is defined in a 3-dimensional (3D) region $\Omega$ as
\begin{equation}
    H^{MHD}[\vb{A}, p, \vb{u}, \rho]=\int_\Omega \left (\frac{1}{2}\rho u^2 +\frac{1}{2}B^2 +\frac{p}{\gamma-1}\right)\sqrt{g} \, \dd^3 x,
    \label{eq:GenMHDhamiltonian}
\end{equation}
where $\vb{A}(\vb{x})$ is the magnetic potential vector, $\vb{B}(\vb{x})=\curl{\vb{A}}$ is the magnetic field, $p(\vb{x})$ is the pressure, $\gamma$ is the adiabatic index, $\sqrt{g}$ is the Jacobian, and $\vb{u}(\vb{x})$ is the generalized flow, defined by $\vb{u} = \boldsymbol{\pi} / \rho$, where $\boldsymbol{\pi}(\vb{x})$ is the momentum density and $\rho (\vb{x})$ is the plasma mass density. \textcolor{blue}{In this work we drop the permeability constant $\mu_0$ from all equations.}

The optimization problem defined as the Task 1 in Ref.~\onlinecite{dewar2022relaxed} is 
\begin{equation}
    \text{Minimize }H^{MHD},\;\;\text{Subject to }\Gamma,
\end{equation}
where $\Gamma$ is the set of all constraints. The conservation of mass, the kinematic constraint of the Lagrangian flow, and the IOL constraint are ``microscopic" constraints; i.e.~ they are applied pointwise to the whole domain. On the other hand, entropy, helicity, and cross helicity constraints are ``macroscopic" constraints; i.e~ \textcolor{blue}{they are defined as integrals taken over the volume $\Omega$ (see \Cref{eq:global_constraint})}. The macroscopic constraints are applied using the Lagrange multiplier method, while the mass conservation and the kinematic constraint are applied explicitly in the variational scheme. The IOL constraint is applied using the augmented Lagrangian method (see Ref.~\onlinecite{nocedal1999numerical}, Chapter 17). After adding the relevant constraints, \Cref{eq:GenMHDhamiltonian} reads\cite{dewar2022relaxed}
    \begin{align}
        H[\vb{A}, p, \Phi, \vb{u}, \vb{v}, \rho, \lambda_s, \lambda_k, \lambda_u] = H^{MHD} +
         \int_\Omega &\left( -\lambda_u \left(\vb{u} \cdot \vb{B} - \frac{U_0}{V_\Omega}\right) 
        - \lambda_k \left(\frac{1}{2} \vb{A} \cdot \vb{B} - \frac{K_0}{V_\Omega}\right) \right.\nonumber \\
           -\lambda_s \left(\frac{\rho}{\gamma - 1} \ln{ \kappa\frac{p}{\rho^\gamma}} - \frac{S_0}{V_\Omega}\right) 
        &\left.-\bm{\lambda}_c \cdot \vb{C} + \frac{1}{2} \mu C^2 \right) \sqrt{g} \, \dd^3 x,
        \label{eq:Genhamiltonianf}
    \end{align}
 where $U_0$, $K_0$, and $S_0$ are the value of the total cross-helicity, helicity, and entropy, respectively, and $\lambda_u$, $\lambda_k$, and $\lambda_s$ are their corresponding Lagrange multipliers (which are independent of position). \textcolor{blue}{$V_\Omega\equiv \int_\Omega \sqrt{g}\dd ^3x$ is the total volume of the domain.}
 In the definition of entropy, $\kappa$ is an arbitrary dimensionalizing constant playing the role of density normalization. 
 The electric potential $\Phi (\vb{x})$ is  defined by $\vb{E}=-\grad{\Phi}$, where $\vb{E}(\vb{x})$ is the electric field. $\vb{C}$ is defined as
\begin{equation}
    \vb{C}\equiv\vb{E}+\vb{v}\times\vb{B}.
    \label{eq:IOL_const}
\end{equation}
The necessity of the generalized flow $\vb{u}$ and its distinction from the other flow $\vb{v}$ are explained in Ref.~\onlinecite{dewar2020time}. In short, $\vb{u}$ is an ingredient of the ``phase space Lagrangian" used to avoid inconsistencies arising from applying the cross helicity constraint in the configuration-space-Lagrangian approach, and $\vb{v}$ is the kinematically constrained Lagrangian flow velocity. In the Hamiltonian minimization approach, we will see that $\vb{u}$ will be a field-aligned velocity while $\vb{v}$ will have a cross-field component. Hence, unlike the MRxMHD the cross-field flow can exist without any need to enforce the angular momentum constraint. 

To avoid the presumptive singularities in the solution \cite{dewar2022relaxed}, the IOL is enforced as a ``weak" constraint. \textcolor{blue}{In Ref.~\onlinecite{dewar2022relaxed}, a weak constraint is defined as ``one that is
enforced as the limiting case of a sequence of soft constraints". This definition, also adopted in this work,  distinguishes a weak constraint from a soft (i.e., approximate) constraint and a hard (i.e., exact) constraint. The augmented Lagrangian provides a method for applying the weak form of IOL, iteratively}. In \Cref{eq:Genhamiltonianf}, $\boldsymbol{\lambda}_c(\vb{x})$ (a position-dependent vector Lagrange multiplier) and $\mu$ are ingredients of the augmented Lagrangian method, used to enforce the ideal Ohm's law constraint pointwise. The advantage of using the augmented Lagrangian method over a penalty method is as follows. If $\bm{\lambda}_c = 0$, the last term in \eqref{eq:Genhamiltonianf} is the penalty term.
Given a fixed $\mu$, the Lagrangian can be extremised to obtain all the physical quantities and Lagrange multipliers, which will, however, depend on the value of $\mu$. However, to enforce a strong constraint, one should prescribe a very large $\mu$, leading to ill-conditioning.
The augmented Lagrangian introduces an additional term $\bm{\lambda}_c$ that mitigates the problem of ill-conditioning: one does not need to go to a very high $\mu$ to achieve the same level of constraint. Instead, an iterative procedure will be used to control the level of constraint. Among the important objectives of this work is to demonstrate the practical feasibility and convergence of the augmented Lagrangian method for enforcing the IOL.

Following Ref.~\onlinecite{nocedal1999numerical} (Chapter 17),  we choose $\{\mu^0,\mu^1, \ldots\}$ an increasing sequence of numbers that is specified as an input and update $\boldsymbol{\lambda}_c$ as 
\begin{equation}
\boldsymbol{\lambda}_c^{N+1}=\boldsymbol{\lambda}_c^N-\mu^N \vb{C}^N,\;\;N=1, 2, \ldots,
\label{eq:IOL_update}
\end{equation}
where $N$ is the iteration number. After each of these iterations (which we call ``IOL iterations"), the model comes closer to satisfying the IOL. This means 
\begin{equation}
    \lim_{N\rightarrow \infty} C^N= 0,
\end{equation}
i.e. as $N\rightarrow\infty$ ( or $\mu\rightarrow \infty$)
\begin{equation}
    \vb{E}+\vb{v}\times \vb{B}=0.
    \label{eq:IOL}
\end{equation}
\textcolor{blue}{Therefore, as $\mu^N$ (or $N$) increases, the model approaches a generally singular state in which the IOL is exactly satisfied. Thus, the index $N$ or the penalty parameter $\mu^N$ can be viewed as the control parameter for the regularization method. The vector $\vb{C}$ resembles $\eta \vb{J}$  ($\eta$ is resistivity) in resistive MHD. Since $\vb{C}$ is position-dependent, the degree to which the ideal Ohm's law (IOL) is satisfied in each IOL iteration varies with position. There is no theoretical limit on $N$, and iterations can proceed as long as the numerical method permits. However, because most numerical methods struggle to handle singularities, the iterations can be terminated at a large $N$ if the solver encounters singularities (or stiffness), as conjectured in Ref.~\onlinecite{dewar2022relaxed}. In this state, $\vb{C}$ may become significantly smaller at some points than at others.
} We note that because of the IOL iterations, all the variable fields as well as $\boldsymbol{\lambda}_c$ and $\mu$ depend on the iteration number $N$ (note that in this work $N$ is not an exponent). To simplify the notation, however, we keep the $N$ index implicit in our notation unless it is necessary to show it explicitly.

The recursive relation of \Cref{eq:IOL_update} can be solved as
\begin{equation}
\boldsymbol{\lambda}_c^N=\boldsymbol{\lambda}_c^0-\sum_{N=0}^{\infty}\mu^N\vb{C}^N.
    \label{eq:IOL_update_solve}
\end{equation}
According to the n-th term test of the divergence of infinite series, the necessary condition for $\lambda_c^\infty$ to converge is 
\begin{equation}
    \lim_{N\rightarrow \infty}\mu^N\vb{C}^N=0,
    \label{eq:div_test}
\end{equation}
which means that for an arbitrarily large $N$, $C$ should vanish faster than $1/\mu$.

The variation of $\lambda_s$, $\lambda_u$, and $\lambda_k$ in \Cref{eq:Genhamiltonianf} 
leads to the macroscopic constraint
\begin{align}
    \int_{\Omega} \frac{\rho}{\gamma-1}\ln{(\kappa\frac{p}{\rho^\gamma})}\sqrt{g}\dd ^3x=S_0,\nonumber\\
    \int_{\Omega} \vb{u}\cdot\vb{B}\sqrt{g}\dd ^3x=U_0,\nonumber\\
    \int_{\Omega}  \frac{1}{2}\vb{A}\cdot\vb{B}\sqrt{g} \dd ^3x=K_0.
    \label{eq:global_constraint}
\end{align}
During the (hypothetical) variation of the Lagrangian coordinates, density is microscopically constrained by the conservation of mass
\begin{equation}
    \delta \rho=-\div{\rho\boldsymbol{\zeta}},
    \label{eq:rho_const}
\end{equation}
where $\boldsymbol{\zeta}\equiv \Delta \vb{x}$ is the variation Lagrangian coordinates (i.e~a ``hypothetical" Lagrangian flow). Also, the Lagrangian flow $\vb{v}$ is constrained by the kinematic constraint in its static form ($\pdv{t}=0$)\cite{newcomb1962lagrangian}, i.e.
\begin{equation}
    \delta \vb{v}= \vb{v}\cdot\nabla \boldsymbol{\zeta}-\boldsymbol{\zeta}\cdot \nabla\vb{v}.
    \label{eq:v_const}
\end{equation}
The constraints on the variations of $\rho$ and $\vb{v}$ are used intrinsically, in deriving the Euler-Lagrange (EL) equations.

\section{Euler Lagrange equations in their 3-dimensional form}
\textcolor{blue}{As discussed in \Cref{sec:setup}, in this work we are focusing on finding the stationary points of the Hamiltonian functional \Cref{eq:Genhamiltonianf}, as opposed to the stationary point of the action integral as done in Ref.~\onlinecite{dewar2022relaxed}.  Because of this, our EL equations are time-independent and slightly simpler than the generic EL equations derived from the phase space Lagrangian in the mentioned study.}

The stationary points of \Cref{eq:Genhamiltonianf} satisfy
\begin{subequations}
\begin{align}
\rho\vb{u}&=\lambda_u\vb{B},\label{eq:uvarEL}\\
    p&=\lambda_s\rho,\label{eq:pvarEL}\\
    \grad{\left(\frac{u^2}{2}-\frac{\lambda_s}{\gamma-1}\ln{\kappa\frac{p}{\rho^\gamma}}\right)}&=\vb{v}\cdot \nabla \vb{w}+\vb{w}\cdot \nabla \vb{v},\label{eq:rhovarEL}\\
    \lambda_k\vb{B}+\lambda_u\curl{\vb{u}}-\curl{(\vb{v}\times(\boldsymbol{\lambda}_c-\mu\vb{C}))}&=\curl{\vb{B}},\label{eq:GenBelt}\\
    \div{(\boldsymbol{\lambda}_c-\mu\vb{C})}&=0,
\end{align}
\label{eq:GenELs}
\end{subequations}
where $\vb{w}\equiv (\boldsymbol{\lambda}_c-\mu \vb{C})\times\vb{B}/\rho$. \Cref{eq:uvarEL,eq:pvarEL} are the results of the variation of $\vb{u}$ and $p$ and are the same as Eqs.~(B7) and (B9) in Ref.~\onlinecite{dewar2020time}, respectively. We note that $\div{\vb{B}}=0$ is implied by \Cref{eq:GenBelt}. From \Cref{eq:pvarEL} one can see that $\lambda_s$ in this model plays the role of the temperature. \Cref{eq:rhovarEL} is a result of the variation of the Lagrangian coordinates, which due to \Cref{eq:rho_const,eq:v_const} results in the variation of both $\rho$ and $\vb{v}$. Eqs.~\ref{eq:GenELs} consists of 11 equations, which are to be solved for 12 scalar fields; i.e.~$\{p,\rho,\Phi\}$ and the components of $\{\vb{B},\vb{u},\vb{v}\}$. \textcolor{blue}{However, one more equation is required to close the system of equations.  This equation is the steady state continuity equation $\div{\rho\vb{v}}=0$, used also in in the derivation of \Cref{eq:rhovarEL} (the continuity equation is not an EL equation). In general, $\vb{v}$ is among the variable fields and should be found as a part of the solution of the system of equations (Eqs.~\ref{eq:GenELs} and $\div{\rho\vb{v}}=0$).} Nevertheless, as we will show later (see \Cref{eq:vchoice}), the simplicity of the geometry chosen in this study allows us to prescribe $\vb{v}$ so that $\div{\rho\vb{v}}=0$ and two of the equations of \Cref{eq:rhovarEL} are trivial. Therefore, with a careful choice of $\vb{v}$, we will eliminate three variable fields and three equations. 

After some reductions Eqs.~\ref{eq:GenELs} read
\begin{subequations}
\begin{align}
    \grad{\qty(\lambda_s \ln \frac{\rho}{\rho_\Omega}+\frac{\lambda_u^2B^2}{2\rho^2})}&=\vb{v}\cdot \nabla \vb{w}+\vb{w}\cdot \nabla \vb{v},\label{eq:GenBer}\\
\lambda_k\vb{B}-\curl{(\vb{v}\times(\boldsymbol{\lambda}_c-\mu\vb{C}))}&=\curl{\qty(1-\frac{\lambda_u^2}{\rho})\vb{B}},\label{eq:GenBelt2}\\
\div{(\boldsymbol{\lambda}_c-\mu\vb{C})}&=0,\label{eq:phivarEL}
\end{align}
\end{subequations}
where $\rho_\Omega\sim \kappa^{1/\gamma-1}$ is an arbitrary constant. \Cref{eq:GenBelt2} is a version of the Beltrami equation modified by the generalized flow $\vb{u}$ and the IOL constraint. This equation can be solved along with the modified Bernoulli's equation (\Cref{eq:GenBer}) and \Cref{eq:phivarEL}. Therefore, the system of \Cref{eq:GenELs} is reduced to \Cref{eq:GenBer,eq:GenBelt2,eq:phivarEL}, and the variable fields in those equations are reduced to $\vb{E}$, $\vb{B}$, and $\rho$. 

If $\div{\vb{v}}=0$ then \Cref{eq:GenBelt2,eq:phivarEL} read
\begin{gather}
       \curl{\qty(1-\frac{\lambda_u^2}{\rho})\vb{B}}=\lambda_k\vb{B}+(\vb{v}\cdot \nabla)\qty(\boldsymbol{\lambda}_c-\mu \vb{E}-\mu\vb{v}\times \vb{B})
        \nonumber\\
        -\qty(\boldsymbol{\lambda}_c-\mu \vb{E}-\mu\vb{v}\times \vb{B})\cdot \grad{\vb{v}}.\label{eq:homv1}
\end{gather}
Also, if we choose $\boldsymbol{\lambda}_c^0$ ($\boldsymbol{\lambda}_c$ at the initial step of optimization) so that $\div{\boldsymbol{\lambda}_c^0}=0$, then comparing \Cref{eq:phivarEL,eq:IOL_update} we have
\begin{equation}
    \div{\boldsymbol{\lambda}_c^N}=0;\;\; N=0,1,2,\ldots\;.
    \label{eq:divlambda}
\end{equation}
Therefore, \Cref{eq:phivarEL} reads $\div{\vb{C}}=0$ or
\begin{equation}
    \div{\vb{E}}=-\div{(\vb{v}\times\vb{B})}.\label{eq:homv2}
\end{equation}
After solving the system of equations, the electric current $\vb{J}$ can be calculated from the Ampers' law $\vb{J}=\curl{\vb{B}}$. 

\textcolor{blue}{The quasi-neutrality requires $\frac{\epsilon_0m_i\abs{\div{E}}}{e\rho}\ll 1$, where $\epsilon_0$ is the electric permittivity of vacuum, $e$ is the electron charge, and $m_i$ is the ion mass. Using \Cref{eq:homv2} one can show that the quasi-neutrality condition leads to $v_\perp\ll\frac{eL_e\rho}{m_i\epsilon_0B}= \tau_A \omega_{pi}c$, where $v_\perp$ is the norm of the cross-field flow, $L_e$ is the characteristic length of inhomogeneity of the electric field, $\tau_A\equiv \sqrt{\rho}L_e/B$ is the Alfv\'en time, $\omega_{pi}\equiv \sqrt{\frac{\rho e^2}{\epsilon_0m_i^2}}$ is the ion plasma frequency, and $c\equiv 1/\sqrt{\epsilon_0}$ is the speed of light. For a fusion grade plasma with $B=5$ T, $\rho=4\times 10^{-7}\;\text{kg}/\text{m}^3$, $L_e=1$ m, and $\omega_{pi}=9\times 10^9$ 1/s, the upper bound $\tau_A \omega_{pi}c$ is much larger than $c$ and therefore essentially non-restrictive. This means, for a fusion grade plasma the model can hardly violate quasi-neutrality even for a very large $v_\perp$.}

\subsection{Implications of the ideal Ohm's law and comparison with nested flux surfaces\label{sec:IOLimplications}}
In Kruskal-Kulsrud and multi-region relaxed (MRxMHD) equilibriums, the nested flux surfaces are supported by $\vb{B}\cdot\grad{p}=0$ and a monotonic radial pressure profile which is provided as an input. For the static equilibrium with a flow,  $\vb{B}\cdot\grad{p}$ does not necessarily vanish. In our case, \Cref{eq:uvarEL,eq:pvarEL} imply
\begin{equation}
    \vb{B}\cdot \grad{p}=-\frac{\lambda_s}{\lambda_u}\rho^2\div{\vb{u}},
\end{equation}
which is not generally zero. However, if IOL is satisfied in a subdomain $\Theta \subseteq \Omega$ with a finite measure, we have
\begin{equation}
    \vb{B}\cdot\grad{\Phi}=0.
    \label{eq:Bgradphi}
\end{equation}
From a Hamiltonian mechanical viewpoint, \Cref{eq:Bgradphi} means that $\Phi$ is a constant of motion,  and if it is differentiable and non-constant then magnetic field lines are tangential to a continuous set of nested flux surfaces in $\Theta$. Therefore,  a ``fully relaxed" region is defined as a region with finite measure inside $\Theta$ where $\grad{\Phi}=0$, as discussed in Refs.~\onlinecite{dewar2020time,dewar2022relaxed}.  In this sense, $\Phi$ plays the same role as $p$ in an equilibrium without flow. However, the implications of $\vb{B}\cdot \grad{\Phi}=0$ and $\vb{B}\cdot \grad{p}=0$ differ because in contrast to the pressure profile, the $\Phi$ profile must be a continuous function, otherwise the $E\cross B$ drift defined by $\frac{-\grad{\Phi}\times\vb{B}}{B^2}$ contracts $\delta$-functions. \textcolor{blue}{A $\delta$-function in the $E\times B$ drift velocity leads to a $\delta^2$-function in the kinetic energy density, which does not have a well-defined integral and therefore is non-physical.}

\textcolor{blue}{It is important to note the definition of a fully relaxed region in this model fundamentally differs from that of Taylor relaxed regions in MRxMHD. In this model, the organization of a fully relaxed region is governed by $\Phi$, which is an output. Consequently, fully relaxed regions with $ \grad{\Phi} = 0$ are self-organized within this framework, rather than being externally imposed as in MRxMHD. This distinction arises from the enforcement of the IOL constraint in this model, which is absent in MRxMHD.
}

 \section{Hamiltonian extremization in a 1D slab geometry}

 In a slab geometry, $x$ resembles the radial direction, $y$ resembles the poloidal direction, and $z$ resembles the azimuthal direction of a torus.  Our slab has a length $2$ that is elongated from $x=-1$ to $x=1$. Using the gauge freedom, we can assume $\vb{A}=\bar{\vb{A}}+\grad{g}$ so that
\begin{subequations}
\begin{align}
    A_x(x,y,z)=\bar{A_x}(x,y,z)+\pdv{g}{x}\;(x,y,z)&=0,\label{eq:gauge1}\\
    A_y(x_+,y,z)=\bar{A_y}(x_+,y,z)+\pdv{g}{y}\,(x_+,y,z)&=\psi_z,\label{eq:gauge2}\\
    A_z(x_+,0,z)=\bar{A_z}(x_+,0,z)+\pdv{g}{z}\,(x_+,0,z)&=-\psi_y,\label{eq:gauge3}
\end{align}
\end{subequations}
where \textcolor{blue}{$x_+$ is the $x$ coordinate of the right boundary}. $\psi_y$ and $\psi_z$ are two given constants that play the role of poloidal and toroidal fluxes, respectively. This choice of gauge is similar to what is used in SPEC\cite{hudson2012computation}. \textcolor{blue}{We note that \Cref{eq:gauge2,eq:gauge3} serve as boundary conditions of $\vb{A}$ and enforce the flux constraints (see \Cref{eq:1DslabBou}).}

We simplify the system of \Cref{eq:GenELs} by assuming that for any scalar $f$, $\pdv{f}{z}=\pdv{f}{y}=0$ (the 1D assumption). This assumption allows us to prescribe $\vb{v}$. In this work, we assume a divergence-free $\vb{v}$ given by
\begin{align}
    \vb{v}=v_z\hat{z}\equiv \omega x\hat{z},
    \label{eq:vchoice}
\end{align}
where $\omega$ is a specified constant. The choice of $\vb{v}$ is inspired by rigid plasma rotation in a torus \cite{finn1983turbulent,dewar2020time}\textcolor{blue}{This is because, in our slab, $x$ and resembles the radial coordinate and $\hat{z}$ resembles the toroidal direction}. We note that \Cref{eq:vchoice} satisfies the continuity equation $\div{\rho \vb{v}}=0$, so long as $\pdv{\rho}{z}=0$.

\subsection{Two models of the optimization}
\begin{figure}
    \centering
    \tikzset{every picture/.style={line width=0.75pt}} 

\begin{tikzpicture}[x=0.75pt,y=0.75pt,yscale=-1,xscale=1]

\draw   (80,19) -- (192.33,19) -- (192.33,56.67) -- (80,56.67) -- cycle ;
\draw    (194,37) -- (225.33,37.63) ;
\draw [shift={(227.33,37.67)}, rotate = 181.15] [color={rgb, 255:red, 0; green, 0; blue, 0 }  ][line width=0.75]    (10.93,-3.29) .. controls (6.95,-1.4) and (3.31,-0.3) .. (0,0) .. controls (3.31,0.3) and (6.95,1.4) .. (10.93,3.29)   ;
\draw   (381,19) -- (496.83,19) -- (496.83,56.67) -- (381,56.67) -- cycle ;
\draw    (347,37) -- (378.33,37.63) ;
\draw [shift={(380.33,37.67)}, rotate = 181.15] [color={rgb, 255:red, 0; green, 0; blue, 0 }  ][line width=0.75]    (10.93,-3.29) .. controls (6.95,-1.4) and (3.31,-0.3) .. (0,0) .. controls (3.31,0.3) and (6.95,1.4) .. (10.93,3.29)   ;
\draw   (79.83,103) -- (193.33,103) -- (193.33,140.67) -- (79.83,140.67) -- cycle ;
\draw    (194,120) -- (225.33,120.63) ;
\draw [shift={(227.33,120.67)}, rotate = 181.15] [color={rgb, 255:red, 0; green, 0; blue, 0 }  ][line width=0.75]    (10.93,-3.29) .. controls (6.95,-1.4) and (3.31,-0.3) .. (0,0) .. controls (3.31,0.3) and (6.95,1.4) .. (10.93,3.29)   ;
\draw   (230,87.67) -- (347.33,87.67) -- (347.33,154.67) -- (230,154.67) -- cycle ;
\draw   (382,103) -- (494.33,103) -- (494.33,140.67) -- (382,140.67) -- cycle ;
\draw    (348,120) -- (379.33,120.63) ;
\draw [shift={(381.33,120.67)}, rotate = 181.15] [color={rgb, 255:red, 0; green, 0; blue, 0 }  ][line width=0.75]    (10.93,-3.29) .. controls (6.95,-1.4) and (3.31,-0.3) .. (0,0) .. controls (3.31,0.3) and (6.95,1.4) .. (10.93,3.29)   ;
\draw   (228,7) -- (345.33,7) -- (345.33,74) -- (228,74) -- cycle ;

\draw (11,28) node [anchor=north west][inner sep=0.75pt]   [align=left] {Model A: \ };
\draw (87,27.4) node [anchor=north west][inner sep=0.75pt]    {$\mathnormal{\lambda _{s} ,\lambda _{u} ,\lambda _{k}} ,\ etc$};
\draw (383,28.4) node [anchor=north west][inner sep=0.75pt]    {$\mathnormal{S_{0} ,\ U_{0} ,\ K_{0}} ,\ etc$};
\draw (11,113) node [anchor=north west][inner sep=0.75pt]   [align=left] {Model B: \ };
\draw (388,111.4) node [anchor=north west][inner sep=0.75pt]    {$\mathnormal{\lambda _{s} ,\lambda _{u} ,\lambda _{k}} ,\ etc$};
\draw (80,112.4) node [anchor=north west][inner sep=0.75pt]    {$\mathnormal{S_0 ,\ U_{0} ,\ K_{0}} ,\ etc$};
\draw (237,89.67) node [anchor=north west][inner sep=0.75]  [font=\small] [align=left] {Minimize $\displaystyle H^{MHD}$\\[-5pt] subject to all\\[-5pt] constraints of $\Gamma$};
\draw (235,11) node [anchor=north west][inner sep=0.75pt]  [font=\small] [align=left] {Minimize $\displaystyle H^{MHD}$\\[-5pt] subject to all\\[-5pt] constraints of $\Gamma$};

\end{tikzpicture}
    \caption{Two different models of the optimization problem.}
    \label{fig:opt_models_fig}
\end{figure}
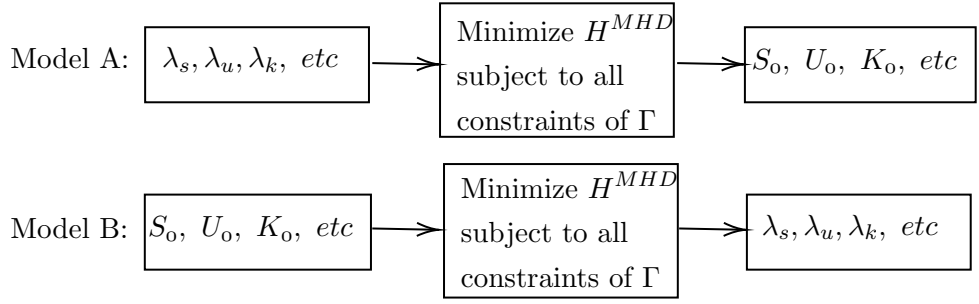

Regarding the macroscopic constraints, there are two ways in the literature that such optimization problems are solved. One might assume that the Lagrange multipliers $\lambda_s$, $\lambda_u$, and $\lambda_k$ are known, solve the Euler-Lagrangian equations, and calculate the $S_0$, $U_0$, and $K_0$ in \Cref{eq:global_constraint} from the solutions. Conversely, given the $S_0$, $U_0$, and $K_0$, one might find the Lagrange multipliers by solving \Cref{eq:global_constraint} in conjunction with the Euler-Lagrange equations. \Cref{fig:opt_models_fig} shows these two models (named model A and B, respectively) schematically.
The technically correct model between these two is model B because it is the only one that satisfies the constraints of \Cref{eq:global_constraint}. However, model A has some advantages that make it helpful.  This model is much simpler to solve because it does not require solving the highly nonlinear equations of \Cref{eq:global_constraint}, for finding ${\lambda_k,\lambda_u,\lambda_s}$. This allows us to solve the optimization problem much more easily and even show an approximate analytical solution to the Euler-Lagrange equations, which we use to verify our numerical solution. Also, later when we numerically solve model B, the Lagrange multipliers used in model A provide one of the solutions of model B, provided that we feed the same $S_0$, $U_0$, and $K_0$ to model B as we have found by solving model A. For this reason, we start solving model A and postpone solving model B to section \ref{sec:modelB}.

\subsection{Hamiltonian functional and EL equations in 1D}
In the slab geometry, \Cref{eq:GenMHDhamiltonian} reads
\begin{gather}
    H[\vb{A},p,\Phi,\vb{u},\rho]=\nonumber\\
    \int_{-1}^1 \left (\frac{1}{2}\rho u^2 +\frac{1}{2}B^2 +\frac{p}{\gamma-1}-\lambda_u \vb{u}\cdot\vb{B}-\lambda_k\frac{1}{2}\vb{A}\cdot\vb{B}-\lambda_s \frac{\rho}{\gamma-1}\ln{\frac{p}{\rho^\gamma}}-\lambda_{cx}(x)C_x(x)+\frac{1}{2}\mu C^2 \right)\dd x,
    \label{eq:hamiltonianf}
\end{gather}
where $\vb{C}=C_x \hat{x}=C\hat{x}$.

Although $\boldsymbol{\lambda}_c$ is generally a vector field, in 1D the $\{y,z\}$ components of $\vb{C}$ vanish and  $\lambda_{cx}(x)\equiv \lambda_c $ appears as a scalar. While in general $\boldsymbol{\lambda}$ and $\vb{C}$ are position dependent, in 1D \Cref{eq:divlambda} imples that both $\lambda_c$ and $C$ are independent of position.
Also in 1D, the $\div{\vb{B}}=0$ implies $B_x$ is constant and this constant should be zero to satisfy $\vb{B}\cdot\hat{n}=0$ at the boundaries. Therefore
\begin{gather}
\vb{B}=B_y\hat{y}+B_z\hat{z}=\partial_xA_y\hat{z}-\partial_xA_z\hat{y}.
\end{gather}
From this expression of $\vb{B}$ and \Cref{eq:vchoice} one can see that, in a 1D slab, the right-hand-side of \Cref{eq:rhovarEL} vanishes. In this case, \Cref{eq:uvarEL,eq:pvarEL,eq:GenBer} imply
\begin{equation}
    p=\lambda_s\rho=\rho_\Omega e^{-u^2/2\lambda_s}.
\end{equation}
For a small parallel Mach number ($M_{\parallel}^2\equiv u^2/\lambda_s\ll 1$), the pressure is approximated as
\begin{equation}
    p\approx \lambda_s\rho_{\Omega}-\lambda_s\rho_{\Omega}\frac{u^2}{2\lambda_s}+ \ldots.
    \label{eq:rhoexpnsion}
\end{equation}
Although we do not use this expansion for constructing a numerical solution, it shows that $\rho=\rho_{\Omega}$ (corresponding to p=$\lambda_s\rho_\Omega$) is a suitable choice of initial density in our iterative numerical method. \Cref{eq:rhoexpnsion} is also used for an approximate analytical solution of the EL equations. 

\textcolor{blue}{By imposing the gauge conditions of \Cref{eq:gauge2,eq:gauge3} and using the remaining gauge freedom to set  $A_y(-1) = A_z(-1) = 0$, one can show}
\begin{align}
    \int_{-1}^1B_z\dd x=A_y(1)&=\psi _z,\nonumber\\
    \int_{-1}^1B_y\dd x=-A_z(1)&=\psi_y,\label{eq:1DslabBou}
\end{align}
which are the boundary conditions of $\vb{A}$.
Therefore, in a 1D slab \Cref{eq:GenBer,eq:homv1,eq:homv2} read
\begin{align}
    \lambda_k\vb{B}
        -\omega\qty(\lambda_c-\mu E_x+\mu \omega x B_y)\hat{z}&=\curl{\qty(1-\frac{\lambda_u^2}{\rho})\vb{B}},\nonumber\\
\dv{E_x}{x}-\omega\dv{\qty(xB_y)}{x}&=0,\nonumber\\
    \lambda_s \ln \frac{\rho}{\rho_\Omega}+\frac{\lambda_u^2B^2}{2\rho^2}&=0.
\end{align}
Separating the $\{y,z\}$ components, using $\vb{B}=\curl{\vb{A}}$ and $\vb{E}=-\grad{\Phi}$,
\begin{subequations}
\begin{align}
    -\lambda_k\dv{A_z}{x}&=\dv{x}\qty[\qty(\frac{\lambda_u^2}{\rho}-1)\dv{A_y}{x}]\label{eq:1DslabBelt1}\\
        \lambda_k\dv{A_y}{x}-\omega\qty(\lambda_c+\mu \dv{\Phi}{x}-\mu \omega x \dv{A_z}{x})&=\dv{x}\qty[\qty(\frac{\lambda_u^2}{\rho}-1)\dv{A_z}{x}]\label{eq:1DslabBelt2}\\
        \dv[2]{\Phi}{x}&=\omega \dv{\qty(x\partial_xA_z)}{x}\label{eq:1DslabE}\\
     -\ln \frac{M_{\parallel}^2}{M_{\parallel \Omega}^2}+M_{\parallel}^2&=0\label{eq:1DslabBer},
\end{align}
\label{eqs:1Dslab}
\end{subequations}
where $M_{\parallel}^2= \frac{\lambda_u^2B^2}{\rho ^2 \lambda_s}=\frac{u^2}{\lambda_s}$ is the parallel Mach number and $M_{\parallel \Omega}^2\equiv \frac{\lambda_u^2B^2}{\rho_\Omega ^2 \lambda_s}$. \Cref{eq:1DslabBelt1,eq:1DslabBelt2} amount to a modified Beltrami equation. For solving \Cref{eq:1DslabE}, we need the potential difference $\Delta \Phi= \Phi (1)-\Phi(-1)$ as an input. A close look at \Cref{eq:1DslabBer} shows that if $M_{\parallel \Omega}^2\le e^{-1}$, it has two real solutions for $M_{\parallel}^2$, otherwise it has no real solution. In the former scenario, one of the solutions is subsonic ($0<M_{\parallel}^2 \le 1$) and the other is supersonic $M_{\parallel}^2\ge 1$. Here, we are only seeking the subsonic solution. 

The IOL iterations are proceeded by \Cref{eq:IOL_update}, which in a 1D slab reads
\begin{equation}
    \lambda_c^{N+1}=\lambda_c^N-\mu^N C^N,
    \label{eq:IOL_update_1D}
\end{equation}
where $C^N=-\dv{\Phi^N}{x}+\omega x \dv{A^{N,z}}{x}$ and $\lambda_c^N$ remain independent of $x$ for $N=0,1,\ldots$. In this work we choose 
\begin{equation}
    \mu^N=\mu^0+N\Delta \mu,
\end{equation}
where $\Delta \mu$ is a specified number. Using this $\mu$, \Cref{eq:IOL_update_solve} reads
\begin{equation}
    \lambda_c^\infty=\lambda_c^0-\mu^0\sum_{N=0}^{\infty}C^N-\Delta \mu\sum_{N=0}^{\infty}NC^N.
    \label{eq:IOL_update_solve_1D}
\end{equation}
In practice, the infinite series are truncated after $N=N_{max}$.

\subsection{Numerical solution of the nonlinear EL equations}

The system of \Cref{eqs:1Dslab} should be solved within each IOL iteration. To this end, we have used the splitting method, which splits the system of \ref{eqs:1Dslab} into two operators. Starting with the initial guess $\rho=\rho_\Omega$, the \Cref{eq:1DslabBelt1,eq:1DslabBelt2,eq:1DslabE} are solved using the boundary values of \Cref{eq:1DslabBou}. The solution provides $\vb{A}$ and $\Phi$, which are then used in solving \Cref{eq:1DslabBer} for $\rho$. Wolfram Mathematica\cite{Mathematica} is used for all numerical analysis in this work. \textcolor{blue}{For solving the differential equations the {\it NDSolve} function with  $InterpolationOrder \rightarrow 4$ option is used \cite{MathematicaNDSolve}.}

\Cref{tab:params} lists the numerical values of the input parameters used for the numerical solution. Among them, $\psi_y$, $\psi_z$, $\lambda_k$ have been tuned so that, if \Cref{eq:1DslabBelt1,eq:1DslabBelt2} are reduced to a Beltrami equation (i.e.~$\omega=\lambda_u=0$), the rotational transform profile reads $\frac{B_y}{B_z}=\tan{\lambda_kx}$ and $\frac{B_y}{B_z}(x=1)=\frac{1+\sqrt{5}}{2}$. The $\lambda_u$ and $\omega$ are chosen as relatively small compared to $\lambda_s$ to resemble the low-Mach-number and high-temperature conditions of stellarator plasmas. In the theoretical formulation of the problem, both $\kappa$ and $\rho_\Omega$ are arbitrary parameters; we set them to 1. 
The exact choice of the $\lambda_c^0$ and the sequence of $\mu$ \textcolor{blue}{is not specified by the theory and} remains essentially arbitrary.  After some trial and error, we came up with a sequence of $\mu={10,20,30,...,200}$ for updating the $\lambda_c^N$ values according to \Cref{eq:IOL_update_1D}.

\begin{figure}[htbp]
    \centering
    \includegraphics[width=\textwidth]{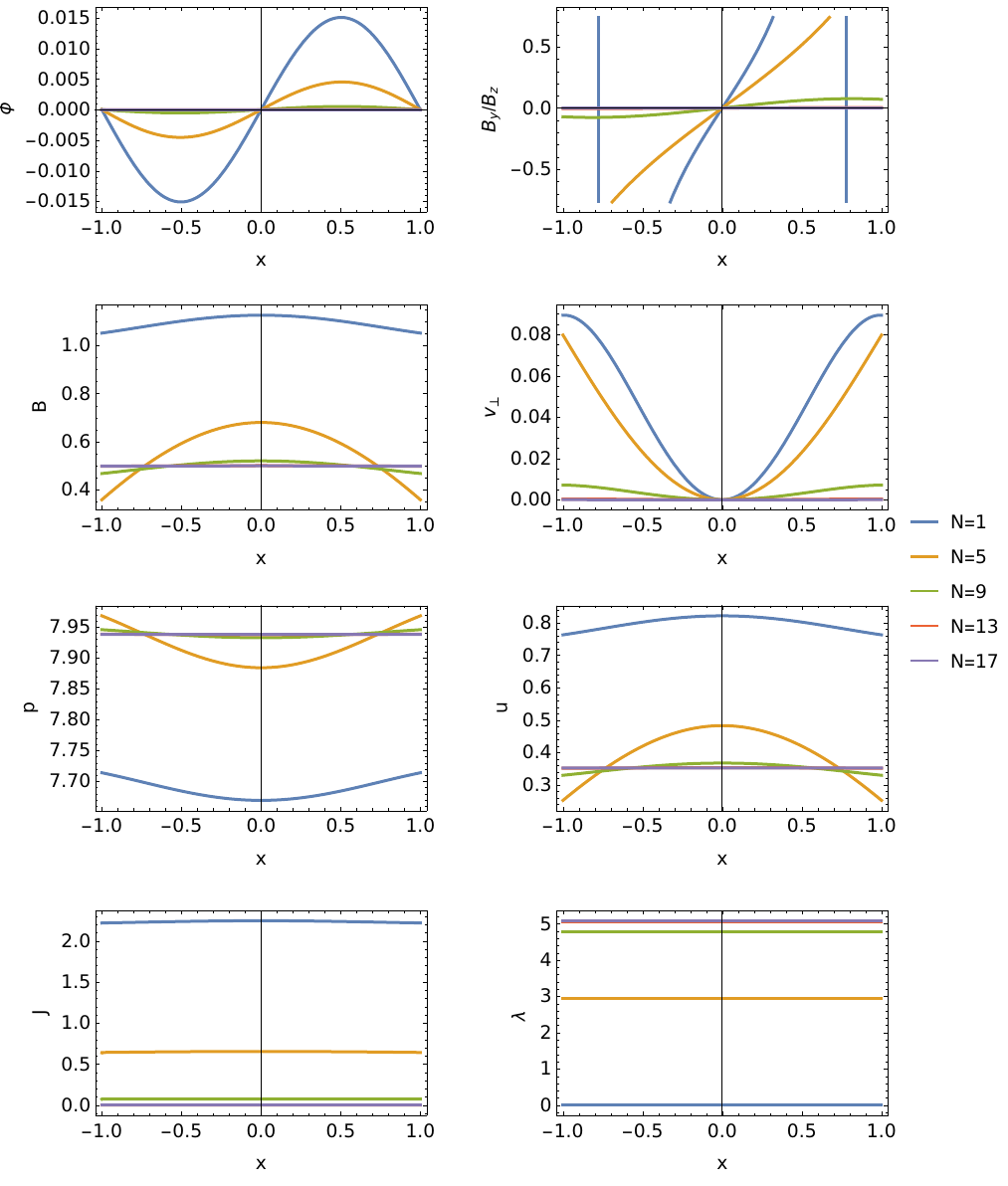}
    \caption{Solution of \Cref{eqs:1Dslab} for several values of $\mu$, as the IOL iterations proceed. $\Delta \Phi=0$.}
    \label{fig:fields_phi0p0_lau0p7}
\end{figure}

\begin{figure}[htbp]
    \centering
    \includegraphics[width=\textwidth]{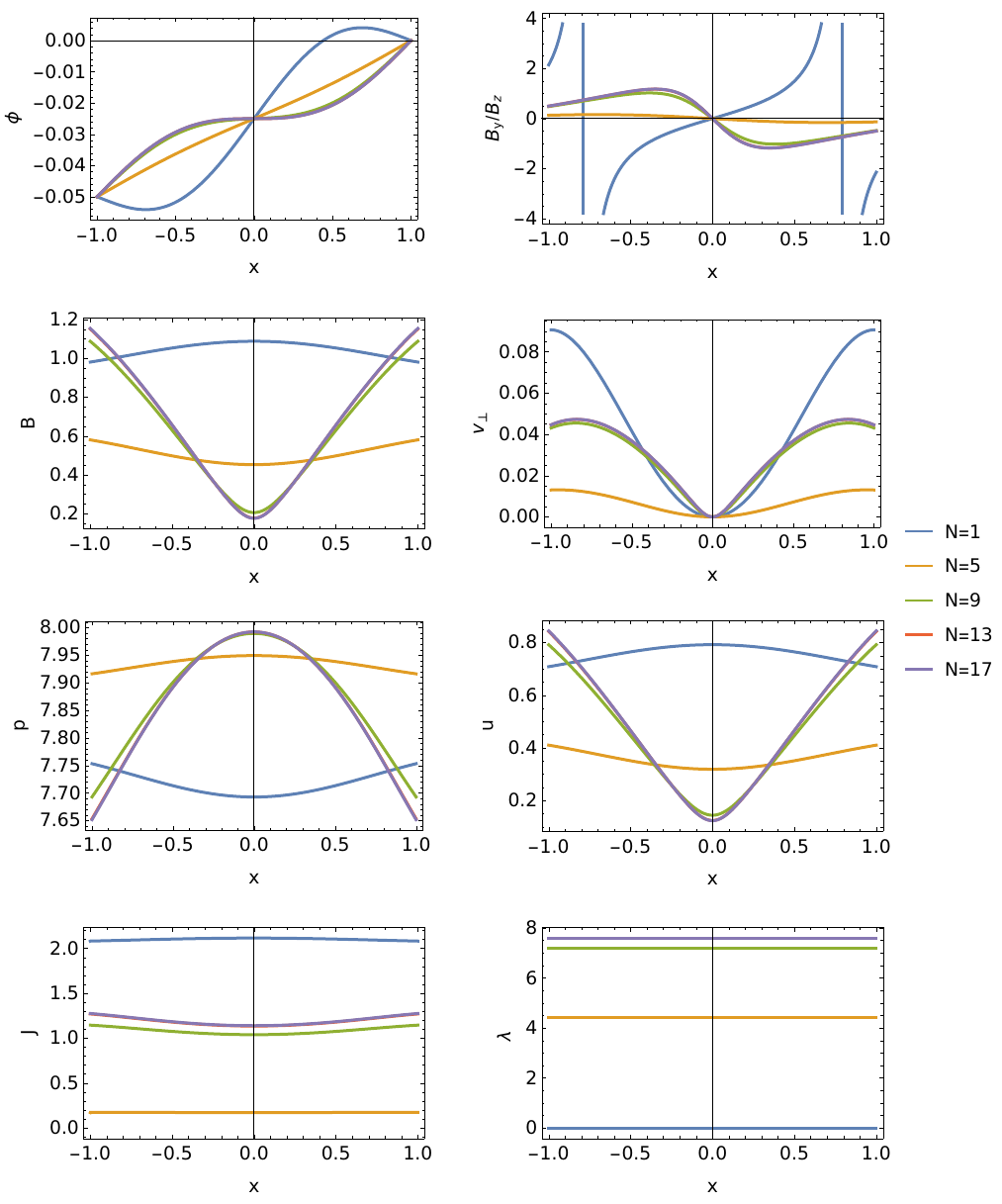}
    \caption{Solution of \Cref{eqs:1Dslab}for several values of $\mu$, as the IOL iterations proceed. $\Delta \Phi=0.05$.}
    \label{fig:fields_phi0p0.05_lau0p7}
\end{figure}

\Cref{fig:fields_phi0p0_lau0p7} shows the numerical solution of different quantities, for the case of $\Delta \Phi=0$. One can see the various stages of the solution's evolution during the IOL iterations, up to $N=17$, in which the IOL is well satisfied. In this state, the electrostatic potential relaxes to $\Phi=0$; this state corresponds to the fully relaxed state as discussed in \Cref{sec:IOLimplications}. The rotational transform profile and the component of $\vb{v}$ perpendicular to $\vb{B}$ ($\vb{v}_\perp\equiv \vb{v}-\vb{v}\cdot\frac{\vb{B}}{B}$) also relax to zero. This is expected from \Cref{eq:IOL} because when $\grad{\Phi}=0$, the $\vb{B}$ is enforced to align with the prescribed $\vb{v}=\omega x\hat{z}$. \Cref{fig:fields_phi0p0.05_lau0p7} shows the solution for a case where $\Delta \Phi \neq 0$ ($\Delta \Phi = 0.05$). In this case, the solution does not converge to the fully relaxed state, and except at $x=0$, $\grad{\Phi}$ remains finite. As a result, a finite $\vb{v}_\perp$ appears. Unlike the MRxMHD, this perpendicular flow does not require any restriction on the geometry of boundaries, and therefore, can potentially exist in a 3D equilibrium.  In every case, one can see that the pressure rises in the positions where the flow $\vb{u}$ drops, as expected for the Bernoulli fluids. According to \Cref{eq:rhoexpnsion}, the pressure gradient is very small, because in this work we have only considered the pressure (and density) changes resulting from a subsonic flow. This characteristic of $p$ can also be observed in \Cref{fig:fields_phi0p0_lau0p7,fig:fields_phi0p0.05_lau0p7}. To prescribe a monotonic pressure profile similar to modern fusion devices, the model should be expanded to incorporate multiple regions, where pressure jumps between regions are supported by magnetic flux surfaces. This is left for future work.

\begin{table}[htbp]
\begin{tabular}{|c|c||c|c|}
\hline
\multicolumn{1}{|c|}{\begin{tabular}[c]{@{}c@{}}Input \\ parameter\end{tabular}} & \multicolumn{1}{c||}{Value(s)}   & \multicolumn{1}{c|}{\begin{tabular}[c]{@{}c@{}}Input\\  parameter\end{tabular}} & \multicolumn{1}{c|}{Value(s)} \\ \hline
$\lambda _s$                                                                     & 8                               & $\kappa$                                                                        & 1                             \\ \hline
$\lambda _u$                                                                     & 0.7                        & $\gamma$                                                                        & 5/3                           \\ \hline
$\lambda _k$                                                                     & $\arctan(\frac{1+\sqrt{5}}{2})$ & $\omega$                                                                        & 0.1                           \\ \hline
$\psi _y$                                                                        & 0                               & $\lambda^0$                                                                     & 0                             \\ \hline
$\psi _z$                                                                        & 1                               & $\mu^0$                                                                         & 10                            \\ \hline
$\Delta \phi$                                                                    & -0.05 to 0.05                     & $\Delta \mu$                                                                    & 10                            \\ \hline
$\rho _\Omega$                                                                   & 1                               & $N_{max}$                                                                       & 20                            \\ \hline
             
\end{tabular}
\caption{The input parameters used for numerical solutions (for model A).}
\label{tab:params}
\end{table}

\subsection{An approximate analytical solution of model A for the small Mach number case}
\begin{figure}[htbp]
    \centering
    \includegraphics[width=\linewidth]{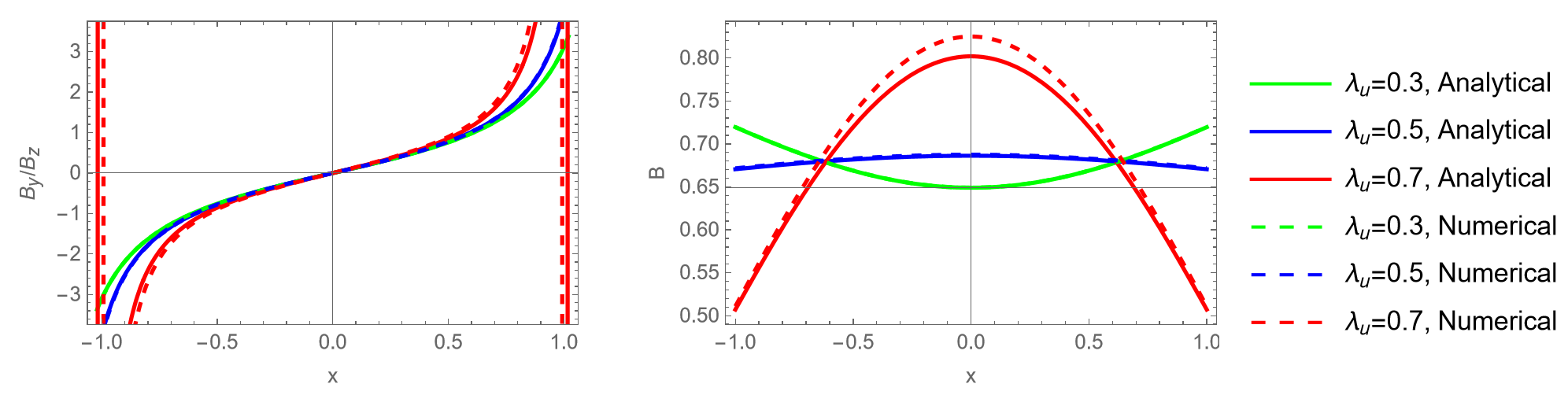}
    \caption{The comparison of the analytical solution (\Cref{eq:an_By,eq:an_Bz}) with the numerical solutions (\Cref{eq:1DslabBelt1,eq:1DslabBelt2,eq:1DslabE,eq:1DslabBer}). $\Delta \Phi=-0.05$.}
    \label{fig:Ana_Num_comp_phi_n0p05}
\end{figure}

\Cref{eqs:1Dslab} can be linearized for $M_{\parallel}^2\ll 1$ and solved analytically, up to a constant and after the IOL is well satisfied. For $M_{\parallel}^2\ll 1$ we approximate $\rho$ with the first term in \Cref{eq:rhoexpnsion} $\rho_{\Omega}$. Using this approximation and assuming that \Cref{eq:div_test} is  satisfied one can show that $N\rightarrow\infty$ \Cref{eq:1DslabBelt1,eq:1DslabBelt2} read
\begin{subequations}
\begin{align}
    -\lambda_k\dv{A_z}{x}&=\qty(\frac{\lambda_u^2}{\rho_\Omega}-1)\dv[2]{A_y}{x},\label{eq:1DslabBelt1_ana}\\
        \lambda_k\dv{A_y}{x}-\omega\lambda_c^\infty&=\qty(\frac{\lambda_u^2}{\rho_\Omega}-1)\dv[2]{A_z}{x}.\label{eq:1DslabBelt2_ana}
\end{align}
\end{subequations}
These equations can be solved analytically and the solution to them gives
\begin{subequations}
\begin{align}
B_y(x)=\frac{\mathcal{K}\csc(\mathcal{K})}{2\lambda_k^2}\qty[\lambda_k^2\psi_y\cos(\mathcal{K}x)-\lambda_k(-2\omega\lambda_c^\infty+\lambda_k\psi_z)\sin(\mathcal{K}x)],\label{eq:an_By}\\
    B_z(x)=\frac{\omega\lambda_c^\infty}{\lambda_k}+\frac{\mathcal{K}\csc(\mathcal{K})}{2\lambda_k^2}\qty[\lambda_k^2\psi_y\sin(\mathcal{K}x)+\lambda_k(-2\omega\lambda_c^\infty+\lambda_k\psi_z)\cos(\mathcal{K}x)],\label{eq:an_Bz}
\end{align}
\end{subequations}
where we have introduced a new constant $\mathcal{K}\equiv \frac{\lambda_k}{(\lambda_u^2/\rho_\Omega-1)}$. Although we calculate $\lambda_c^\infty$ in \Cref{eq:an_Bz} using \Cref{eq:IOL_update_1D} and the numerical solutions of $C^N$, we shall call \Cref{eq:an_By,eq:an_Bz} a semi-analytical solution, or for brevity ``analytical solution". \Cref{fig:Ana_Num_comp_phi_n0p05} compares the semi-analytical and numerical solutions for different values of $\lambda_u$. In this figure, $B_y/B_z$ is the rotational transform and $B$ is the norm of the magnetic field. The parameters used for this figure are again what is listed in \Cref{tab:params}. For $\lambda_u\leq 0.5$, one can see an excellent agreement between the analytical and numerical solutions, which validates our numerical results.
For $\lambda_u=0.7$, singularities can be seen in the rotational transition profile close to boundaries where $B_z$ vanishes. Using \Cref{eq:an_Bz}, one can calculate that $B_z$ vanishes at about $x\approx \pm 1.017$. In the numerical solution, this point slightly moves to $x\approx \pm 1$.

\section{Model B of the optimization: Optimization for fixed entropy, cross helicity, and helicity}\label{sec:modelB}
So far in developing the numerical and analytical solutions, we have used model A; i.e~we have assumed that $\lambda_s$, $\lambda_u$, and $\lambda_k$ are specified. However, in the complete optimization model (called model B) we should variate these Lagrange multipliers to keep entropy, cross helicity and helicity fixed. To this end, one needs to solve Eqs.~\ref{eq:global_constraint} along with Eqs.~\ref{eqs:1Dslab}. Because Eqs.~\ref{eq:global_constraint} are highly nonlinear, model B leads to a highly nonlinear problem that is more challenging to solve. Analytical solutions are barely possible in this model, and the computational cost for a numerical solution is significantly increased.

The method we use in the model B of optimization is similar to the shooting method.  In this method, the three algebraic Eqs.~\ref{eq:global_constraint} are solved for $\lambda_s$, $\lambda_u$, and $\lambda_k$ using the Newton method. Because these algebraic equations depend on the variable fields $\vb{A}$, $\vb{u}$, $\vb{\rho}$, and $\vb{p}$, within each iteration of the Newton method, we need to solve the system of differential equations of \Cref{eqs:1Dslab} and provide these variable fields. The initial guess of $\lambda_s$, $\lambda_u$, and $\lambda_k$ for the Newton method is always provided by the latest results, except for the first IOL iteration where these values are arbitrarily guessed. Like model A, $\Delta \mu=10$ is used in model B. The choice of the value of increment of the penalty parameter $\Delta \mu$ can significantly impact the computational cost of the model. A very large $\Delta \mu$ can lead to difficulties in the convergence of the Newton method, while a very small $\Delta \mu$ necessitates a larger $N_{max}$ which can lead to an increased computational cost. \textcolor{blue}{As to the choice of $\mu_0$, we observed that a value of zero leads to the stagnation of the augmented Lagrangian in the initial step, while a value $\mu_0\gg \Delta \mu$ leads to the convergence problems in the Newton method. Therefore, $\mu_0=\Delta \mu=10$ was chosen.}

\begin{figure}[htbp]
    \centering
    \includegraphics[width=0.49\textwidth]{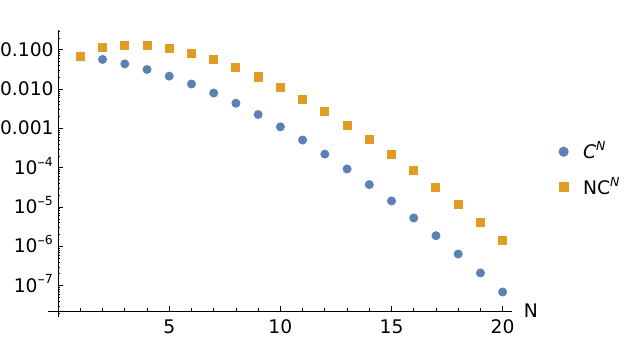}
    \includegraphics[width=0.49\textwidth]{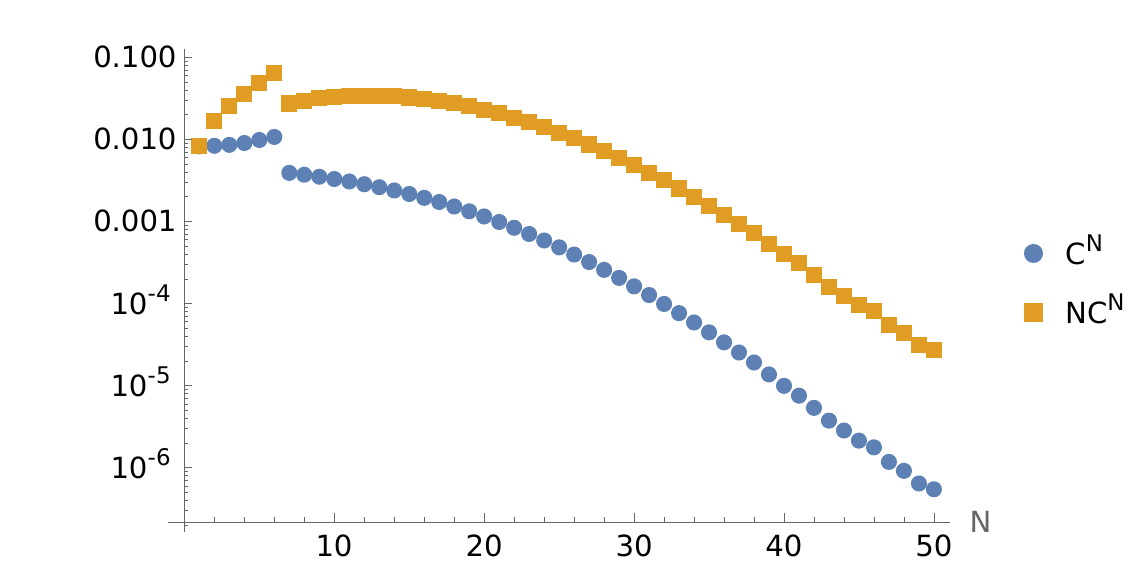}
    \caption{The convergence of $C$ and $NC$ during IOL iterations. Left) Model A, Right) Model B; $\Delta \Phi=0.05$.}  
    \label{fig:CN_vs_N}
\end{figure}

The report on our numerical results is as follows. In model A, the set of Lagrange multipliers $\{\lambda_s=8,\lambda_u=0.7,\lambda_k=1.01722\}$ is mapped to  $\{S_0=6.17,U_0=0.75,K_0=-0.18\}$ after 20 IOL iterations. In model B, the same values of $\{S_0,U_0,K_0\}$ are mapped to $\{\lambda_s=8.00714,\lambda_u=0.70052,\lambda_k=1.02735\}$ after 50 IOL iterations. Therefore, the resultant values of Lagrange multipliers in model B are the same as the input values of model A, within an error of less than 1 percent. The convergence of the augmented Lagrangian method in both models A and B is shown in \Cref{fig:CN_vs_N}. In both cases, the convergence is exponential for the large N and the condition $\lim_{N\rightarrow\infty} NC^N=0$ required for the convergence of series in \Cref{eq:IOL_update_solve_1D}. However, the convergence of Model A is smoother and faster than that of Model B. \textcolor{blue}{In the right panel of \Cref{fig:CN_vs_N}, $C^N$ initially increases but then decreases after a few IOL iterations, resulting in a sharp transition in the convergence plot. This behaviour stems from the specific choice of the $\mu$ sequence; however, it is inconsequential as long as the results for $N \gg 1$ remain accurate.}

\textcolor{blue}{On an \texttt{Intel\textsuperscript{\textregistered} Core\texttrademark{} i7-14700} (5.4~GHz, 20~cores/28~threads) and using only 1 core, the IOL iterations took 1 minute for Model A, while they took about 4 hours for Model B. One might need to use parallel computing for more complicated cases, especially for the Model B. It is possible to parallelize the Newton method used in Model B inside each IOL iteration; e.g.,~in constructing the Jacobian matrix (a task that involves solving the EL equations) or in solving the linear system of equations created by the Newton method. Another approach for boosting the performance of model B is designing and using a preconditioner that can improve the convergence of the Newton method.}

\section{Conclusion and discussion}
The numerical feasibility and convergence of the RxMHD with IOL constraints are demonstrated through a 1-dimensional slab model. The validity of the numerical results is demonstrated by comparing them with analytical solutions. This study is the first step toward an ``extended MRxMHD model".

RxMHD with IOL constraint provides a mathematical framework for the relaxation of the ideal MHD while closely conforming to it. This model can enhance the existing MRxMHD theory by the natural inclusion of the cross-field flow. \textcolor{blue}{The model however does not include any non-ideal or kinetic effect such as the neoclassical transport or bootstrap current. As a result, the compatibility with such theories can only be found {\it a posteriori}.} Although we have not considered the time-dependent case, it has been considered in the original theory. Thus, unlike in MRxMHD, the path to developing a time-dependent code is clear in this model. Unlike SPEC which relies on an input number of relaxed regions, in this model the fully relaxed regions (defined by $\grad{\Phi}=0$) can be self-organized through a mechanism discussed in \Cref{sec:IOLimplications} and demonstrated numerically in \Cref{fig:fields_phi0p0_lau0p7}. However, due to assumed geometrical symmetries (the 1D assumption) magnetic islands and chaos cannot be formed in fully relaxed regions. \textcolor{blue}{Another potential extension of this work is to incorporate multiple regions, allowing for pressure profiles similar to those in fusion devices. Similar to MRxMHD, this model would consist of an arbitrary number of regions separated by finite pressure jumps and ideal interfaces, but these regions would not necessarily be relaxed in the MRxMHD sense. Depending on the resulting $\Phi$, a region could be fully relaxed, non-relaxed, or a combination of both. Furthermore, one can try to position ideal interfaces in the non-relaxed parts where they are naturally expected. Eventually, as conjectured \cite{DewarPrivateComm}, this approach might help mitigate the non-existence issue of MRxMHD. However, this conjecture remains to be tested in future studies.} Applying this model to a 3-dimensional geometry and including multiple regions are left for future works.

\section*{Acknowledgment}
We dedicate this study to Prof. R. L. Dewar, the founder of the theoretical model discussed in this work, whose pioneering contributions have profoundly shaped the development of relaxed magnetohydrodynamics. This research was supported by a grant from the Simons Foundation/SFARI (560651, AB), and the U.S. Department of Energy under contract number DE-AC02-09CH11466. The United States Government retains a non-exclusive, paid-up, irrevocable, world-wide license to publish or reproduce the published form of this manuscript, or allow others to do so, for United States Government purposes. 

 \section*{Author Declarations}
 \subsection*{Conflict of interest}
 The authors have no conflicts to disclose.
 
 \section*{Data Availability Statement}
 The data that support the findings of this study are available from the corresponding author
upon reasonable request.

\bibliographystyle{apsrev4-2}
\bibliography{Refs.bib}

\begin{thebibliography}{34}%
\makeatletter
\providecommand \@ifxundefined [1]{%
 \@ifx{#1\undefined}
}%
\providecommand \@ifnum [1]{%
 \ifnum #1\expandafter \@firstoftwo
 \else \expandafter \@secondoftwo
 \fi
}%
\providecommand \@ifx [1]{%
 \ifx #1\expandafter \@firstoftwo
 \else \expandafter \@secondoftwo
 \fi
}%
\providecommand \natexlab [1]{#1}%
\providecommand \enquote  [1]{``#1''}%
\providecommand \bibnamefont  [1]{#1}%
\providecommand \bibfnamefont [1]{#1}%
\providecommand \citenamefont [1]{#1}%
\providecommand \href@noop [0]{\@secondoftwo}%
\providecommand \href [0]{\begingroup \@sanitize@url \@href}%
\providecommand \@href[1]{\@@startlink{#1}\@@href}%
\providecommand \@@href[1]{\endgroup#1\@@endlink}%
\providecommand \@sanitize@url [0]{\catcode `\\12\catcode `\$12\catcode
  `\&12\catcode `\#12\catcode `\^12\catcode `\_12\catcode `\%12\relax}%
\providecommand \@@startlink[1]{}%
\providecommand \@@endlink[0]{}%
\providecommand \url  [0]{\begingroup\@sanitize@url \@url }%
\providecommand \@url [1]{\endgroup\@href {#1}{\urlprefix }}%
\providecommand \urlprefix  [0]{URL }%
\providecommand \Eprint [0]{\href }%
\providecommand \doibase [0]{https://doi.org/}%
\providecommand \selectlanguage [0]{\@gobble}%
\providecommand \bibinfo  [0]{\@secondoftwo}%
\providecommand \bibfield  [0]{\@secondoftwo}%
\providecommand \translation [1]{[#1]}%
\providecommand \BibitemOpen [0]{}%
\providecommand \bibitemStop [0]{}%
\providecommand \bibitemNoStop [0]{.\EOS\space}%
\providecommand \EOS [0]{\spacefactor3000\relax}%
\providecommand \BibitemShut  [1]{\csname bibitem#1\endcsname}%
\let\auto@bib@innerbib\@empty
\bibitem [{\citenamefont {Kruskal}\ and\ \citenamefont
  {Kulsrud}(1958)}]{kruskal1958equilibrium}%
  \BibitemOpen
  \bibfield  {author} {\bibinfo {author} {\bibfnamefont {M.~D.}\ \bibnamefont
  {Kruskal}}\ and\ \bibinfo {author} {\bibfnamefont {R.}~\bibnamefont
  {Kulsrud}},\ }\href@noop {} {\emph {\bibinfo {title} {Equilibrium of a
  magnetically confined plasma in a toroid}}}\ (\bibinfo  {publisher}
  {Princeton University Plasma Physics Laboratory.},\ \bibinfo {year}
  {1958})\BibitemShut {NoStop}%
\bibitem [{\citenamefont {Lichtenberg}\ and\ \citenamefont
  {Lieberman}(2013)}]{lichtenberg2013regular}%
  \BibitemOpen
  \bibfield  {author} {\bibinfo {author} {\bibfnamefont {A.~J.}\ \bibnamefont
  {Lichtenberg}}\ and\ \bibinfo {author} {\bibfnamefont {M.~A.}\ \bibnamefont
  {Lieberman}},\ }\href@noop {} {\emph {\bibinfo {title} {Regular and chaotic
  dynamics}}},\ Vol.~\bibinfo {volume} {38}\ (\bibinfo  {publisher} {Springer
  Science \& Business Media},\ \bibinfo {year} {2013})\BibitemShut {NoStop}%
\bibitem [{\citenamefont {Rodr{\'\i}guez}\ and\ \citenamefont
  {Bhattacharjee}(2021)}]{rodriguez2021islands}%
  \BibitemOpen
  \bibfield  {author} {\bibinfo {author} {\bibfnamefont {E.}~\bibnamefont
  {Rodr{\'\i}guez}}\ and\ \bibinfo {author} {\bibfnamefont {A.}~\bibnamefont
  {Bhattacharjee}},\ }\href@noop {} {\bibfield  {journal} {\bibinfo  {journal}
  {Physics of Plasmas}\ }\textbf {\bibinfo {volume} {28}} (\bibinfo {year}
  {2021})}\BibitemShut {NoStop}%
\bibitem [{\citenamefont {Hudson}\ and\ \citenamefont
  {Nakajima}(2010)}]{hudson2010pressure}%
  \BibitemOpen
  \bibfield  {author} {\bibinfo {author} {\bibfnamefont {S.}~\bibnamefont
  {Hudson}}\ and\ \bibinfo {author} {\bibfnamefont {N.}~\bibnamefont
  {Nakajima}},\ }\href@noop {} {\bibfield  {journal} {\bibinfo  {journal}
  {Physics of Plasmas}\ }\textbf {\bibinfo {volume} {17}} (\bibinfo {year}
  {2010})}\BibitemShut {NoStop}%
\bibitem [{\citenamefont {Taylor}(1986)}]{taylor1986relaxation}%
  \BibitemOpen
  \bibfield  {author} {\bibinfo {author} {\bibfnamefont {J.~B.}\ \bibnamefont
  {Taylor}},\ }\href@noop {} {\bibfield  {journal} {\bibinfo  {journal}
  {Reviews of Modern Physics}\ }\textbf {\bibinfo {volume} {58}},\ \bibinfo
  {pages} {741} (\bibinfo {year} {1986})}\BibitemShut {NoStop}%
\bibitem [{\citenamefont {Taylor}(1974)}]{taylor1974relaxation}%
  \BibitemOpen
  \bibfield  {author} {\bibinfo {author} {\bibfnamefont {J.~B.}\ \bibnamefont
  {Taylor}},\ }\href@noop {} {\bibfield  {journal} {\bibinfo  {journal}
  {Physical Review Letters}\ }\textbf {\bibinfo {volume} {33}},\ \bibinfo
  {pages} {1139} (\bibinfo {year} {1974})}\BibitemShut {NoStop}%
\bibitem [{\citenamefont {Bodin}(1990)}]{bodin1990reversed}%
  \BibitemOpen
  \bibfield  {author} {\bibinfo {author} {\bibfnamefont {H.}~\bibnamefont
  {Bodin}},\ }\href@noop {} {\bibfield  {journal} {\bibinfo  {journal} {Nuclear
  Fusion}\ }\textbf {\bibinfo {volume} {30}},\ \bibinfo {pages} {1717}
  (\bibinfo {year} {1990})}\BibitemShut {NoStop}%
\bibitem [{\citenamefont {Yamada}\ \emph {et~al.}(2010)\citenamefont {Yamada},
  \citenamefont {Kulsrud},\ and\ \citenamefont {Ji}}]{yamada2010magnetic}%
  \BibitemOpen
  \bibfield  {author} {\bibinfo {author} {\bibfnamefont {M.}~\bibnamefont
  {Yamada}}, \bibinfo {author} {\bibfnamefont {R.}~\bibnamefont {Kulsrud}},\
  and\ \bibinfo {author} {\bibfnamefont {H.}~\bibnamefont {Ji}},\ }\href@noop
  {} {\bibfield  {journal} {\bibinfo  {journal} {Reviews of modern physics}\
  }\textbf {\bibinfo {volume} {82}},\ \bibinfo {pages} {603} (\bibinfo {year}
  {2010})}\BibitemShut {NoStop}%
\bibitem [{\citenamefont {Hole}\ \emph {et~al.}(2006)\citenamefont {Hole},
  \citenamefont {Hudson},\ and\ \citenamefont {Dewar}}]{hole2006stepped}%
  \BibitemOpen
  \bibfield  {author} {\bibinfo {author} {\bibfnamefont {M.}~\bibnamefont
  {Hole}}, \bibinfo {author} {\bibfnamefont {S.~R.}\ \bibnamefont {Hudson}},\
  and\ \bibinfo {author} {\bibfnamefont {R.}~\bibnamefont {Dewar}},\
  }\href@noop {} {\bibfield  {journal} {\bibinfo  {journal} {Journal of Plasma
  Physics}\ }\textbf {\bibinfo {volume} {72}},\ \bibinfo {pages} {1167}
  (\bibinfo {year} {2006})}\BibitemShut {NoStop}%
\bibitem [{\citenamefont {Hudson}\ \emph {et~al.}(2012)\citenamefont {Hudson},
  \citenamefont {Dewar}, \citenamefont {Dennis}, \citenamefont {Hole},
  \citenamefont {McGann}, \citenamefont {Von~Nessi},\ and\ \citenamefont
  {Lazerson}}]{hudson2012computation}%
  \BibitemOpen
  \bibfield  {author} {\bibinfo {author} {\bibfnamefont {S.}~\bibnamefont
  {Hudson}}, \bibinfo {author} {\bibfnamefont {R.}~\bibnamefont {Dewar}},
  \bibinfo {author} {\bibfnamefont {G.}~\bibnamefont {Dennis}}, \bibinfo
  {author} {\bibfnamefont {M.}~\bibnamefont {Hole}}, \bibinfo {author}
  {\bibfnamefont {M.}~\bibnamefont {McGann}}, \bibinfo {author} {\bibfnamefont
  {G.}~\bibnamefont {Von~Nessi}},\ and\ \bibinfo {author} {\bibfnamefont
  {S.}~\bibnamefont {Lazerson}},\ }\href@noop {} {\bibfield  {journal}
  {\bibinfo  {journal} {Physics of Plasmas}\ }\textbf {\bibinfo {volume} {19}}
  (\bibinfo {year} {2012})}\BibitemShut {NoStop}%
\bibitem [{\citenamefont {Dewar}\ \emph {et~al.}(2008)\citenamefont {Dewar},
  \citenamefont {Hole}, \citenamefont {McGann}, \citenamefont {Mills},\ and\
  \citenamefont {Hudson}}]{dewar2008relaxed}%
  \BibitemOpen
  \bibfield  {author} {\bibinfo {author} {\bibfnamefont {R.~L.}\ \bibnamefont
  {Dewar}}, \bibinfo {author} {\bibfnamefont {M.~J.}\ \bibnamefont {Hole}},
  \bibinfo {author} {\bibfnamefont {M.}~\bibnamefont {McGann}}, \bibinfo
  {author} {\bibfnamefont {R.}~\bibnamefont {Mills}},\ and\ \bibinfo {author}
  {\bibfnamefont {S.~R.}\ \bibnamefont {Hudson}},\ }\href@noop {} {\bibfield
  {journal} {\bibinfo  {journal} {Entropy}\ }\textbf {\bibinfo {volume} {10}},\
  \bibinfo {pages} {621} (\bibinfo {year} {2008})}\BibitemShut {NoStop}%
\bibitem [{\citenamefont {Loizu}\ \emph {et~al.}(2020)\citenamefont {Loizu},
  \citenamefont {Huang}, \citenamefont {Hudson}, \citenamefont {Baillod},
  \citenamefont {Kumar},\ and\ \citenamefont {Qu}}]{loizu2020direct}%
  \BibitemOpen
  \bibfield  {author} {\bibinfo {author} {\bibfnamefont {J.}~\bibnamefont
  {Loizu}}, \bibinfo {author} {\bibfnamefont {Y.-M.}\ \bibnamefont {Huang}},
  \bibinfo {author} {\bibfnamefont {S.}~\bibnamefont {Hudson}}, \bibinfo
  {author} {\bibfnamefont {A.}~\bibnamefont {Baillod}}, \bibinfo {author}
  {\bibfnamefont {A.}~\bibnamefont {Kumar}},\ and\ \bibinfo {author}
  {\bibfnamefont {Z.}~\bibnamefont {Qu}},\ }\href@noop {} {\bibfield  {journal}
  {\bibinfo  {journal} {Physics of Plasmas}\ }\textbf {\bibinfo {volume} {27}}
  (\bibinfo {year} {2020})}\BibitemShut {NoStop}%
\bibitem [{\citenamefont {Dennis}\ \emph {et~al.}(2013)\citenamefont {Dennis},
  \citenamefont {Hudson}, \citenamefont {Terranova}, \citenamefont {Franz},
  \citenamefont {Dewar},\ and\ \citenamefont {Hole}}]{dennis2013minimally}%
  \BibitemOpen
  \bibfield  {author} {\bibinfo {author} {\bibfnamefont {G.}~\bibnamefont
  {Dennis}}, \bibinfo {author} {\bibfnamefont {S.~R.}\ \bibnamefont {Hudson}},
  \bibinfo {author} {\bibfnamefont {D.}~\bibnamefont {Terranova}}, \bibinfo
  {author} {\bibfnamefont {P.}~\bibnamefont {Franz}}, \bibinfo {author}
  {\bibfnamefont {R.}~\bibnamefont {Dewar}},\ and\ \bibinfo {author}
  {\bibfnamefont {M.}~\bibnamefont {Hole}},\ }\href@noop {} {\bibfield
  {journal} {\bibinfo  {journal} {Physical Review Letters}\ }\textbf {\bibinfo
  {volume} {111}},\ \bibinfo {pages} {055003} (\bibinfo {year}
  {2013})}\BibitemShut {NoStop}%
\bibitem [{\citenamefont {Loizu}\ \emph {et~al.}(2017)\citenamefont {Loizu},
  \citenamefont {Hudson}, \citenamefont {N{\"u}hrenberg}, \citenamefont
  {Geiger},\ and\ \citenamefont {Helander}}]{loizu2017equilibrium}%
  \BibitemOpen
  \bibfield  {author} {\bibinfo {author} {\bibfnamefont {J.}~\bibnamefont
  {Loizu}}, \bibinfo {author} {\bibfnamefont {S.}~\bibnamefont {Hudson}},
  \bibinfo {author} {\bibfnamefont {C.}~\bibnamefont {N{\"u}hrenberg}},
  \bibinfo {author} {\bibfnamefont {J.}~\bibnamefont {Geiger}},\ and\ \bibinfo
  {author} {\bibfnamefont {P.}~\bibnamefont {Helander}},\ }\href@noop {}
  {\bibfield  {journal} {\bibinfo  {journal} {Journal of Plasma Physics}\
  }\textbf {\bibinfo {volume} {83}},\ \bibinfo {pages} {715830601} (\bibinfo
  {year} {2017})}\BibitemShut {NoStop}%
\bibitem [{\citenamefont {Loizu}\ \emph {et~al.}(2016)\citenamefont {Loizu},
  \citenamefont {Hudson},\ and\ \citenamefont
  {N{\"u}hrenberg}}]{loizu2016verification}%
  \BibitemOpen
  \bibfield  {author} {\bibinfo {author} {\bibfnamefont {J.}~\bibnamefont
  {Loizu}}, \bibinfo {author} {\bibfnamefont {S.}~\bibnamefont {Hudson}},\ and\
  \bibinfo {author} {\bibfnamefont {C.}~\bibnamefont {N{\"u}hrenberg}},\
  }\href@noop {} {\bibfield  {journal} {\bibinfo  {journal} {Physics of
  Plasmas}\ }\textbf {\bibinfo {volume} {23}} (\bibinfo {year}
  {2016})}\BibitemShut {NoStop}%
\bibitem [{\citenamefont {Hudson}\ \emph {et~al.}(2020)\citenamefont {Hudson},
  \citenamefont {Loizu}, \citenamefont {Zhu}, \citenamefont {Qu}, \citenamefont
  {N{\"u}hrenberg}, \citenamefont {Lazerson}, \citenamefont {Smiet},\ and\
  \citenamefont {Hole}}]{hudson2020free}%
  \BibitemOpen
  \bibfield  {author} {\bibinfo {author} {\bibfnamefont {S.}~\bibnamefont
  {Hudson}}, \bibinfo {author} {\bibfnamefont {J.}~\bibnamefont {Loizu}},
  \bibinfo {author} {\bibfnamefont {C.}~\bibnamefont {Zhu}}, \bibinfo {author}
  {\bibfnamefont {Z.}~\bibnamefont {Qu}}, \bibinfo {author} {\bibfnamefont
  {C.}~\bibnamefont {N{\"u}hrenberg}}, \bibinfo {author} {\bibfnamefont
  {S.}~\bibnamefont {Lazerson}}, \bibinfo {author} {\bibfnamefont
  {C.}~\bibnamefont {Smiet}},\ and\ \bibinfo {author} {\bibfnamefont
  {M.}~\bibnamefont {Hole}},\ }\href@noop {} {\bibfield  {journal} {\bibinfo
  {journal} {Plasma Physics and Controlled Fusion}\ }\textbf {\bibinfo {volume}
  {62}},\ \bibinfo {pages} {084002} (\bibinfo {year} {2020})}\BibitemShut
  {NoStop}%
\bibitem [{\citenamefont {Loizu}\ and\ \citenamefont
  {Bonfiglio}(2023)}]{loizu2023nonlinear}%
  \BibitemOpen
  \bibfield  {author} {\bibinfo {author} {\bibfnamefont {J.}~\bibnamefont
  {Loizu}}\ and\ \bibinfo {author} {\bibfnamefont {D.}~\bibnamefont
  {Bonfiglio}},\ }\href@noop {} {\bibfield  {journal} {\bibinfo  {journal}
  {Journal of Plasma Physics}\ }\textbf {\bibinfo {volume} {89}},\ \bibinfo
  {pages} {905890507} (\bibinfo {year} {2023})}\BibitemShut {NoStop}%
\bibitem [{\citenamefont {Balkovic}\ \emph {et~al.}(2024)\citenamefont
  {Balkovic}, \citenamefont {Loizu}, \citenamefont {Graves}, \citenamefont
  {Huang},\ and\ \citenamefont {Smiet}}]{balkovic2024direct}%
  \BibitemOpen
  \bibfield  {author} {\bibinfo {author} {\bibfnamefont {E.}~\bibnamefont
  {Balkovic}}, \bibinfo {author} {\bibfnamefont {J.}~\bibnamefont {Loizu}},
  \bibinfo {author} {\bibfnamefont {J.~P.}\ \bibnamefont {Graves}}, \bibinfo
  {author} {\bibfnamefont {Y.-M.}\ \bibnamefont {Huang}},\ and\ \bibinfo
  {author} {\bibfnamefont {C.}~\bibnamefont {Smiet}},\ }\href@noop {}
  {\bibfield  {journal} {\bibinfo  {journal} {Plasma Physics and Controlled
  Fusion}\ }\textbf {\bibinfo {volume} {67}},\ \bibinfo {pages} {015009}
  (\bibinfo {year} {2024})}\BibitemShut {NoStop}%
\bibitem [{\citenamefont {Qu}\ \emph {et~al.}(2020)\citenamefont {Qu},
  \citenamefont {Dewar}, \citenamefont {Ebrahimi}, \citenamefont {Anderson},
  \citenamefont {Hudson},\ and\ \citenamefont {Hole}}]{qu2020stepped}%
  \BibitemOpen
  \bibfield  {author} {\bibinfo {author} {\bibfnamefont {Z.}~\bibnamefont
  {Qu}}, \bibinfo {author} {\bibfnamefont {R.~L.}\ \bibnamefont {Dewar}},
  \bibinfo {author} {\bibfnamefont {F.}~\bibnamefont {Ebrahimi}}, \bibinfo
  {author} {\bibfnamefont {J.~K.}\ \bibnamefont {Anderson}}, \bibinfo {author}
  {\bibfnamefont {S.~R.}\ \bibnamefont {Hudson}},\ and\ \bibinfo {author}
  {\bibfnamefont {M.~J.}\ \bibnamefont {Hole}},\ }\href@noop {} {\bibfield
  {journal} {\bibinfo  {journal} {Plasma Physics and Controlled Fusion}\
  }\textbf {\bibinfo {volume} {62}},\ \bibinfo {pages} {054002} (\bibinfo
  {year} {2020})}\BibitemShut {NoStop}%
\bibitem [{\citenamefont {Dennis}\ \emph {et~al.}(2014)\citenamefont {Dennis},
  \citenamefont {Hudson}, \citenamefont {Dewar},\ and\ \citenamefont
  {Hole}}]{dennis2014multi}%
  \BibitemOpen
  \bibfield  {author} {\bibinfo {author} {\bibfnamefont {G.}~\bibnamefont
  {Dennis}}, \bibinfo {author} {\bibfnamefont {S.}~\bibnamefont {Hudson}},
  \bibinfo {author} {\bibfnamefont {R.}~\bibnamefont {Dewar}},\ and\ \bibinfo
  {author} {\bibfnamefont {M.}~\bibnamefont {Hole}},\ }\href@noop {} {\bibfield
   {journal} {\bibinfo  {journal} {Physics of Plasmas}\ }\textbf {\bibinfo
  {volume} {21}} (\bibinfo {year} {2014})}\BibitemShut {NoStop}%
\bibitem [{\citenamefont {Bruno}\ and\ \citenamefont
  {Laurence}(1996)}]{bruno1996existence}%
  \BibitemOpen
  \bibfield  {author} {\bibinfo {author} {\bibfnamefont {O.~P.}\ \bibnamefont
  {Bruno}}\ and\ \bibinfo {author} {\bibfnamefont {P.}~\bibnamefont
  {Laurence}},\ }\href@noop {} {\bibfield  {journal} {\bibinfo  {journal}
  {Communications on pure and applied mathematics}\ }\textbf {\bibinfo {volume}
  {49}},\ \bibinfo {pages} {717} (\bibinfo {year} {1996})}\BibitemShut
  {NoStop}%
\bibitem [{\citenamefont {Qu}\ \emph {et~al.}(2021)\citenamefont {Qu},
  \citenamefont {Hudson}, \citenamefont {Dewar}, \citenamefont {Loizu},\ and\
  \citenamefont {Hole}}]{qu2021non}%
  \BibitemOpen
  \bibfield  {author} {\bibinfo {author} {\bibfnamefont {Z.~S.}\ \bibnamefont
  {Qu}}, \bibinfo {author} {\bibfnamefont {S.~R.}\ \bibnamefont {Hudson}},
  \bibinfo {author} {\bibfnamefont {R.~L.}\ \bibnamefont {Dewar}}, \bibinfo
  {author} {\bibfnamefont {J.}~\bibnamefont {Loizu}},\ and\ \bibinfo {author}
  {\bibfnamefont {M.~J.}\ \bibnamefont {Hole}},\ }\href@noop {} {\bibfield
  {journal} {\bibinfo  {journal} {Plasma Physics and Controlled Fusion}\
  }\textbf {\bibinfo {volume} {63}},\ \bibinfo {pages} {125007} (\bibinfo
  {year} {2021})}\BibitemShut {NoStop}%
\bibitem [{\citenamefont {Bhattacharjee}\ \emph {et~al.}(1980)\citenamefont
  {Bhattacharjee}, \citenamefont {Dewar},\ and\ \citenamefont
  {Monticello}}]{bhattacharjee1980energy}%
  \BibitemOpen
  \bibfield  {author} {\bibinfo {author} {\bibfnamefont {A.}~\bibnamefont
  {Bhattacharjee}}, \bibinfo {author} {\bibfnamefont {R.}~\bibnamefont
  {Dewar}},\ and\ \bibinfo {author} {\bibfnamefont {D.}~\bibnamefont
  {Monticello}},\ }\href@noop {} {\bibfield  {journal} {\bibinfo  {journal}
  {Physical Review Letters}\ }\textbf {\bibinfo {volume} {45}},\ \bibinfo
  {pages} {347} (\bibinfo {year} {1980})}\BibitemShut {NoStop}%
\bibitem [{\citenamefont {Yeates}\ \emph {et~al.}(2010)\citenamefont {Yeates},
  \citenamefont {Hornig},\ and\ \citenamefont
  {Wilmot-Smith}}]{yeates2010topological}%
  \BibitemOpen
  \bibfield  {author} {\bibinfo {author} {\bibfnamefont {A.}~\bibnamefont
  {Yeates}}, \bibinfo {author} {\bibfnamefont {G.}~\bibnamefont {Hornig}},\
  and\ \bibinfo {author} {\bibfnamefont {A.}~\bibnamefont {Wilmot-Smith}},\
  }\href@noop {} {\bibfield  {journal} {\bibinfo  {journal} {Physical Review
  Letters}\ }\textbf {\bibinfo {volume} {105}},\ \bibinfo {pages} {085002}
  (\bibinfo {year} {2010})}\BibitemShut {NoStop}%
\bibitem [{\citenamefont {Amari}\ and\ \citenamefont
  {Luciani}(2000)}]{amari2000helicity}%
  \BibitemOpen
  \bibfield  {author} {\bibinfo {author} {\bibfnamefont {T.}~\bibnamefont
  {Amari}}\ and\ \bibinfo {author} {\bibfnamefont {J.}~\bibnamefont
  {Luciani}},\ }\href@noop {} {\bibfield  {journal} {\bibinfo  {journal}
  {Physical Review Letters}\ }\textbf {\bibinfo {volume} {84}},\ \bibinfo
  {pages} {1196} (\bibinfo {year} {2000})}\BibitemShut {NoStop}%
\bibitem [{\citenamefont {Dewar}\ \emph {et~al.}(2015)\citenamefont {Dewar},
  \citenamefont {Yoshida}, \citenamefont {Bhattacharjee},\ and\ \citenamefont
  {Hudson}}]{dewar2015variational}%
  \BibitemOpen
  \bibfield  {author} {\bibinfo {author} {\bibfnamefont {R.~L.}\ \bibnamefont
  {Dewar}}, \bibinfo {author} {\bibfnamefont {Z.}~\bibnamefont {Yoshida}},
  \bibinfo {author} {\bibfnamefont {A.}~\bibnamefont {Bhattacharjee}},\ and\
  \bibinfo {author} {\bibfnamefont {S.~R.}\ \bibnamefont {Hudson}},\
  }\href@noop {} {\bibfield  {journal} {\bibinfo  {journal} {Journal of Plasma
  Physics}\ }\textbf {\bibinfo {volume} {81}},\ \bibinfo {pages} {515810604}
  (\bibinfo {year} {2015})}\BibitemShut {NoStop}%
\bibitem [{\citenamefont {Dewar}\ \emph {et~al.}(2020)\citenamefont {Dewar},
  \citenamefont {Burby}, \citenamefont {Qu}, \citenamefont {Sato},\ and\
  \citenamefont {Hole}}]{dewar2020time}%
  \BibitemOpen
  \bibfield  {author} {\bibinfo {author} {\bibfnamefont {R.~L.}\ \bibnamefont
  {Dewar}}, \bibinfo {author} {\bibfnamefont {J.~W.}\ \bibnamefont {Burby}},
  \bibinfo {author} {\bibfnamefont {Z.}~\bibnamefont {Qu}}, \bibinfo {author}
  {\bibfnamefont {N.}~\bibnamefont {Sato}},\ and\ \bibinfo {author}
  {\bibfnamefont {M.}~\bibnamefont {Hole}},\ }\href@noop {} {\bibfield
  {journal} {\bibinfo  {journal} {Physics of Plasmas}\ }\textbf {\bibinfo
  {volume} {27}} (\bibinfo {year} {2020})}\BibitemShut {NoStop}%
\bibitem [{\citenamefont {Dewar}\ and\ \citenamefont
  {Qu}(2022)}]{dewar2022relaxed}%
  \BibitemOpen
  \bibfield  {author} {\bibinfo {author} {\bibfnamefont {R.}~\bibnamefont
  {Dewar}}\ and\ \bibinfo {author} {\bibfnamefont {Z.}~\bibnamefont {Qu}},\
  }\href@noop {} {\bibfield  {journal} {\bibinfo  {journal} {Journal of Plasma
  Physics}\ }\textbf {\bibinfo {volume} {88}},\ \bibinfo {pages} {835880101}
  (\bibinfo {year} {2022})}\BibitemShut {NoStop}%
\bibitem [{\citenamefont {Dewar}(2023)}]{DewarPrivateComm}%
  \BibitemOpen
  \bibfield  {author} {\bibinfo {author} {\bibfnamefont {R.}~\bibnamefont
  {Dewar}},\ }\href@noop {} {\bibinfo {title} {Private communication}}
  (\bibinfo {year} {2023})\BibitemShut {NoStop}%
\bibitem [{\citenamefont {Nocedal}\ and\ \citenamefont
  {Wright}(1999)}]{nocedal1999numerical}%
  \BibitemOpen
  \bibfield  {author} {\bibinfo {author} {\bibfnamefont {J.}~\bibnamefont
  {Nocedal}}\ and\ \bibinfo {author} {\bibfnamefont {S.~J.}\ \bibnamefont
  {Wright}},\ }\href@noop {} {\emph {\bibinfo {title} {Numerical
  optimization}}}\ (\bibinfo  {publisher} {Springer},\ \bibinfo {year}
  {1999})\BibitemShut {NoStop}%
\bibitem [{\citenamefont {Newcomb}(1962)}]{newcomb1962lagrangian}%
  \BibitemOpen
  \bibfield  {author} {\bibinfo {author} {\bibfnamefont {W.}~\bibnamefont
  {Newcomb}},\ }\href@noop {} {\bibfield  {journal} {\bibinfo  {journal}
  {Nuclear Fusion}\ ,\ \bibinfo {pages} {451}} (\bibinfo {year}
  {1962})}\BibitemShut {NoStop}%
\bibitem [{\citenamefont {Finn}\ and\ \citenamefont
  {Antonsen}(1983)}]{finn1983turbulent}%
  \BibitemOpen
  \bibfield  {author} {\bibinfo {author} {\bibfnamefont {J.~M.}\ \bibnamefont
  {Finn}}\ and\ \bibinfo {author} {\bibfnamefont {T.}~\bibnamefont
  {Antonsen}},\ }\href@noop {} {\bibfield  {journal} {\bibinfo  {journal} {The
  Physics of fluids}\ }\textbf {\bibinfo {volume} {26}},\ \bibinfo {pages}
  {3540} (\bibinfo {year} {1983})}\BibitemShut {NoStop}%
\bibitem [{\citenamefont {Inc.}()}]{Mathematica}%
  \BibitemOpen
  \bibfield  {author} {\bibinfo {author} {\bibfnamefont {W.~R.}\ \bibnamefont
  {Inc.}},\ }\href {https://www.wolfram.com/mathematica} {\bibinfo {title}
  {Mathematica, {V}ersion 14.1}},\ \bibinfo {note} {champaign, IL,
  2024}\BibitemShut {NoStop}%
\bibitem [{\citenamefont {Wolfram~Research}(2024)}]{MathematicaNDSolve}%
  \BibitemOpen
  \bibfield  {author} {\bibinfo {author} {\bibfnamefont {I.}~\bibnamefont
  {Wolfram~Research}},\ }\href
  {https://reference.wolfram.com/language/ref/NDSolve.html} {\bibinfo {title}
  {Mathematica, {Version} 14.1}} (\bibinfo {year} {2024}),\ \bibinfo {note}
  {{NDS}olve {D}ocumentation}\BibitemShut {NoStop}%
\end{thebibliography}%

\end{document}